\documentclass[11pt]{article}
\usepackage[a4paper, margin=1.2in]{geometry}
\usepackage[final]{graphicx}
\usepackage[utf8]{inputenc}
\usepackage[english]{babel}
\usepackage{amsmath}
\usepackage{mathtools}
\usepackage{url}

\usepackage{ifpdf}
\ifpdf
\usepackage[%
  pdftitle={},%
  pdfauthor={},%
  pdfstartview=FitH,%
  bookmarks=false,%
  bookmarksopen=false,%
  breaklinks=true,%
  colorlinks=true,%
  linkcolor=blue,anchorcolor=blue,%
  citecolor=blue,filecolor=blue,%
  menucolor=black,pagecolor=blue,%
  urlcolor=blue]{hyperref}
\else
\usepackage[%
  breaklinks=true,%
  colorlinks=true,%
  linkcolor=blue,anchorcolor=blue,%
  citecolor=blue,filecolor=blue,%
  menucolor=blue,pagecolor=blue,%
  urlcolor=blue]{hyperref}
\fi
\usepackage{bm}

\newcommand{\beginsupplement}{
        \renewcommand{\thesubsection}{S.\arabic{subsection}}
     }

\begin{document}

\begin{center}
        \Large{Estimating the course of the COVID-19 pandemic in Germany \\ via spline-based hierarchical modelling of death counts}
\end{center}

\vspace{.2cm}

\begin{center}
{\large
Tobias Wistuba, Andreas Mayr, Christian Staerk\footnote[1]{ \textit{Address for
correspondence:} Dr.~Christian Staerk, Department of Medical Biometry, Informatics and Epidemiology, University Hospital Bonn, Venusberg-Campus 1, 53127 Bonn.\\ Email:
\href{mailto:christian.staerk$at$imbie.uni-bonn.de}{christian.staerk@imbie.uni-bonn.de}} 
}
\end{center}

\begin{center}
\textit{Working Group Statistical Methods in Epidemiology\\ Department of Medical Biometry, Informatics and Epidemiology \\ University Hospital Bonn, Germany} \\
\end{center}

\vspace{.1cm}

\begin{abstract}
    The effective reproduction number is a key figure to monitor the course of the COVID-19 pandemic. Typically, it is estimated based on the day-by-day evolution of confirmed cases. In this study we consider a retrospective modelling approach for estimating the effective reproduction number based on death counts during the first year of the pandemic in Germany. The proposed Bayesian hierarchical model incorporates splines to estimate reproduction numbers flexibly over time while adjusting for varying effective infection fatality rates. The approach also provides estimates of dark figures regarding undetected infections over time. Results for Germany illustrate that estimated reproduction numbers based on death counts are often similar to classical estimates 
    based on confirmed cases. However, considering death counts proves to be more robust against shifts in testing policies: in particular, confirmed cases indicate a short-term spike in the effective reproduction number linked to a local super-spreading event in June~2020, whereas our model does not estimate a spike during this period but reduced dark figures of infections. 
    During the second wave of infections, classical estimation of the reproduction number suggests a flattening/ decreasing trend of infections following the ``lockdown light'' in November~2020, while our results indicate that true numbers of infections continued to rise until the ``second lockdown'' in December~2020. This observation is associated with more stringent testing criteria introduced concurrently with the ``lockdown light'', which is reflected in subsequently increasing dark figures of infections estimated by our model. These findings illustrate that the retrospective viewpoint can provide additional insights regarding the course of the pandemic. In light of progressive vaccinations, shifting the focus from modelling confirmed cases to reported deaths with the possibility to incorporate effective infection fatality rates might be of increasing relevance for the future surveillance of the pandemic.  
    
    \paragraph{Keywords:} COVID-19; SARS-CoV-2; Bayesian Modelling; Effective Reproduction Number; Dark Figures

\end{abstract}
\section{Introduction} \label{sec:intro}

The COVID-19 pandemic continues to have severe impacts on public health in many parts of the world. During the course of the pandemic, different non-pharmaceutical interventions (NPIs) have been implemented to mitigate the spread of the virus, including closures of schools and nurseries, cancellation of public events, regulations regarding social distancing, closures of non-essential shops and further measures~(see e.g.~\cite{ebrahim2020}). Since many of these interventions impose a large burden on society and economy, it is crucial to understand which measures are effective in reducing the spread of the virus. A central figure in this context is the time-dependent effective reproduction number, which can be interpreted as the average number of cases infected by one case in a particular region at a certain time. In practice, the effective reproduction number is usually estimated based on currently available surveillance data, particularly based on daily numbers of newly confirmed cases~(see e.g.~\cite{heiden2020}). While numbers of confirmed cases are crucial for nowcasting the current development of the effective reproduction number \cite{gunther2021, schneble2020}, they are largely influenced by the implemented testing policies. In particular, short-term changes in numbers of conducted SARS-CoV-2 tests complicate the accurate estimation of the effective reproduction number. 

Data on COVID-19 related deaths can provide an additional, retrospective viewpoint on the course of the pandemic and the assessment of NPIs. Important and influential approaches hence also focused on modelling the spread of SARS-CoV-2 based on numbers of reported deaths \cite{flaxman2020estimating, brauner2021}. In particular, the Imperial College COVID-19 Response Team \cite{flaxman2020estimating} developed a Bayesian hierarchical model to estimate the impact of NPIs on the effective reproduction number in different European countries during the first wave of infections in spring~2020.  In the further course of the pandemic, several interventions (e.g., closures of schools and non-essential shops) have been adapted, relaxed or restricted to particular regions, while others (e.g, cancellations of public events and face masks regulations) have largely kept in place. To account for this development, the Bayesian model of Flaxman et al.~\cite{flaxman2020estimating} has been updated and extended to estimate the effectiveness of NPIs during the second infection wave in Europe~\cite{sharma2021}. 

Other authors have argued, that resulting effect estimates of NPIs are non-robust and highly model-dependent \cite{chin2021, wood2021}. In particular, the selection of NPIs to be included in the model predetermine the potential change points for the effective reproduction number and thus can have large effects on the estimates attributed to individual prevention measures. Furthermore, not only the implemented NPIs change over time, but also the adherence and awareness of the population, which may not be adequately described by categorical variables for the implemented prevention measures (cf.~\cite{bonisch2020, schlosser2020} for analyses of mobility patterns during the first phase of the pandemic in Germany). Another limitation of the original model of Flaxman et al.~\cite{flaxman2020estimating} is that the infection fatality rate~(IFR) is assumed to be constant over time. However, the IFR of COVID-19 increases largely with increasing age \cite{o2021, levin2020}. As the age distribution of infections changes substantially during the course of the pandemic~\cite{staerk2021}, the effective IFR should not be regarded as constant over time when modelling the number of infections based on deaths data. 

In this study we adapt the Bayesian hierarchical model of Flaxman et al.~\cite{flaxman2020estimating} to estimate the course of the effective reproduction number in Germany by continuous smoothing splines, without the need for additional information regarding the timings of specific interventions. While our model is primarily driven by the numbers of reported deaths similar as in previous modelling approaches \cite{flaxman2020estimating, wood2021}, we additionally incorporate the changing age distribution of confirmed infections to account for changes in effective IFR. We compare our model estimates of the effective reproduction number with classical estimates derived solely from the numbers of confirmed cases in combination with nowcasting~\cite{heiden2020, rkiRwert2021}. Furthermore, we discuss resulting estimates of dark figures of infections over the course of the pandemic in Germany. Complimentary results for the individual 16 German federal states are provided in the Supplement to this paper (see Section~\ref{ssec:states}).

\section{Methods} \label{sec:methods}

\begin{figure}[t!]
 \centering
 \includegraphics[width=0.9\textwidth]{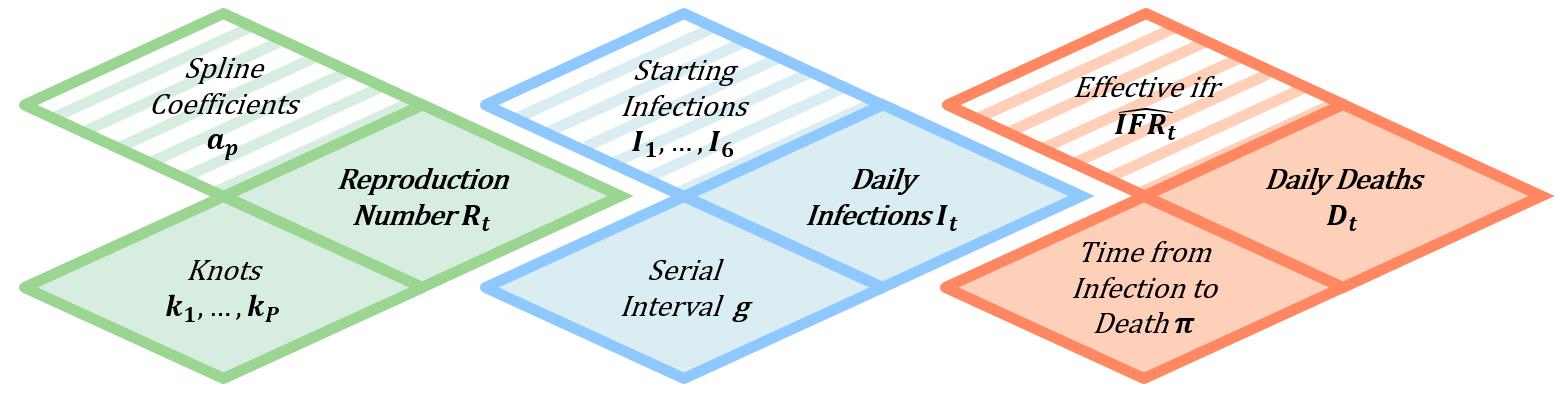}
 \caption{Schematic overview of the adapted Bayesian hierarchical model (cf.~Flaxman et al.~\cite{flaxman2020estimating}).}
 \label{fig:modelflow}
\end{figure}
 
 To estimate the course of the pandemic in Germany based on death counts, we extend the Bayesian hierarchical model from Flaxman et al.~\cite{flaxman2020estimating} by adjusting for time-dependent effective infection fatality rates and by considering splines for modelling the effective reproduction number over time. Figure~\ref{fig:modelflow} provides a schematic representation of the adapted hierarchical model. The main idea of the model is to estimate the effective reproduction number~\(R_t\) retrospectively from daily numbers of reported deaths~\(D_t\) associated with COVID-19. For this, we use death counts for Germany (and also for the 16 German federal states, see Section~\ref{ssec:states} of the Supplement) based on daily situation reports published by the German federal agency for disease control and prevention (Robert Koch Institute, RKI;~\cite{rkisit2021}). 
 
 On the last level of the hierarchical model (see Figure~\ref{fig:modelflow}), the reported deaths~$D_{t}$ are modelled with a negative binomial distribution, i.e.\ the likelihood is given by 
\begin{align}
    \begin{split}
        D_{t} & \sim \text{NB}\Bigl(d_{t},\, d_{t}+\frac{d_{t}^{2}}{\psi}\Bigr)
    \end{split}
    \label{likelihood}
\end{align}
with mean~$d_t$ and variance \(d_{t}+\frac{d_{t}^{2}}{\psi}\). A half-normal distribution is considered as the prior for the dispersion parameter~$\psi\sim \mathcal{N}^{+}(0,\,5)$. 
Numbers of daily reported deaths~\(D_t\) are linked to expected numbers of daily infections~\(I_\tau\), \(\tau<t\), by taking into account estimates for the effective infection fatality rate (IFR) and the time distribution between infection and reported death. 
More specifically, expected numbers of daily reported deaths~$d_{t}$ are given by
\begin{align}
    d_{t} = \sum_{\tau = 1}^{t-1}  \pi_\tau \cdot \widehat{\text{IFR}}_{\tau}   \cdot I_{\tau} ,
\end{align}
where~\(\pi\) denotes the (discretized) distribution for the time between infection and reported death for lethal infections, \(\widehat{\text{IFR}}_{\tau}\) is the estimated effective infection fatality rate and \(I_{\tau}\) denotes the expected number of infections for day~\(\tau\). The distribution~$\pi$ for the time between infection and reported death is obtained as the sum of two components: the incubation period and the time between symptom onset and reported death. 
Based on results from Lauer et al.~\cite{lauer2020incubation}, for the incubation period we use a log-normal distribution with mean~5.52 (days) and standard deviation~2.43 (days). The distribution for the time between symptom onset and reported death is adopted from Flaxman et al.~\cite{flaxman2020estimating} and given by a gamma distribution with mean~17.8 (days) and standard deviation~8.01 (days). Since our model is based on daily data, we consider a discretized version~$\pi$ of the distribution for the sum of the two periods.  

The IFR is an important link when trying to infer total numbers of infections from death counts. As the IFR of COVID-19 largely depends on the age of the infected individuals~\cite{o2021, levin2020, brazeau2020}, it is important to take the age structure of infections into account, which has been shown to vary over the course of the pandemic in Germany~\cite{staerk2021}. Thus, in contrast to the original model of Flaxman et al.~\cite{flaxman2020estimating} using time-constant averaged IFRs, we consider time-dependent effective IFRs which reflect the changing age distribution of infections. Here, we estimate weekly effective IFRs based on the assumption that the age distributions of true infections can be approximated by the age distributions of confirmed cases (cf.~\cite{staerk2021}). In particular, based on data from the RKI~\cite{rkitests}, let~$C_{a,w}$ denote the number of confirmed cases in calendar week~$w$ for age-group $a\in \{$0\scalebox{0.75}[1.0]{\( - \)}9, 10\scalebox{0.75}[1.0]{\( - \)}19, \ldots, 70\scalebox{0.75}[1.0]{\( - \)}79, 80+$\}$. Furthermore, let  $\widehat{\text{IFR}}_{a}$ denote the estimated infection fatality rate for age group~$a$ based on Brazeau et al.~\cite{brazeau2020} (see Section~\ref{ssec:sensitivity} of the Supplement for sensitivity analyses using alternative age-specific IFR estimates from O'Driscoll et al.~\cite{o2021} and Levin et al.~\cite{levin2020}). 
The effective IFR for day~$\tau$ in calendar week~$w$ is estimated as a weighted average of age-specific IFR estimates: 
\begin{align}
    \widehat{\text{IFR}}_{\tau} = \frac{1}{C_{w}} \sum_{a\in A} C_{a,w} \cdot \widehat{\text{IFR}}_{a} .
    \label{ifrf}
\end{align}
Finally, IFR estimates are shifted backwards by ten days to adjust for the delay between the reporting date of cases~\(C_t\) and the date of infections~\(I_t\). This delay can be expressed as the sum of the incubation time with an expected value of 5.52~days~\cite{lauer2020incubation} and the time between onset of symptoms and reporting date (median delay of four days for German data~\cite{covid19data}). 

On the previous level of the hierarchical model (see Figure~\ref{fig:modelflow}), expected numbers of daily infections~\(I_t\) are modelled based on estimated infections~$I_\tau$ of the preceding days~$\tau<t$ via
\begin{align}
    I_{t} = R_{t}\cdot \sum_{\tau=1}^{t-1} I_{\tau}\cdot g_{t-\tau},
    \label{casef}
\end{align}
where~$R_t$ denotes the effective reproduction number and $g$ the (discretized) generation time distribution (i.e.\ the time between two infections in a transmission pair). As infection times are generally unknown, direct estimation of the generation time is rather difficult and commonly approximated by the serial interval.  Based on results from Nishiura et al.~\cite{nishiura2020serial}, the generation time~$g$ is modelled using a discretized log-normal distribution with mean $4.7$~(days) and standard deviation $2.9$~(days). 
An important and idealistic assumption of the considered semi-mechanistic model is that the country is viewed as a closed environment (cf.~\cite{flaxman2020estimating}), so that all infections are assumed to occur within the German population. Consequently, numbers of infections for the first days have to be initialized.  
Similarly to Flaxman et al.~\cite{flaxman2020estimating}, for the first six modelling days $t=1,\dots,6$, expected numbers of infections~\(I_t\sim\text{Exp}(1/\lambda)\) are  exponentially distributed with mean  $\lambda\sim\text{Exp}(10)$.    
The starting date for modelling (\(t=1\)) is considered to be the 15th of January 2020, which is defined as the earliest date so that 60 days later, at least 10 cumulative deaths associated with COVID-19 have been reported in Germany. 

On the first level of the hierarchical model (see Figure~\ref{fig:modelflow}), the effective reproduction number is modelled with a smoothing spline via
\begin{align}
    R_{t} = \max \Bigl( \sum_{p}a_{p} B_{p,3}(t),\,0\Bigr) ,
    \label{rtf}
\end{align}
where $B_{p,3}$ are B-splines of third degree between equidistant knots (each with distance of two weeks) and $a_{p}$ denote the corresponding spline coefficients~(cf.~\cite{kharratzadeh2017splines}). The maximum in equation~\eqref{rtf} is taken to ensure that the effective reproduction number is non-negative. 
For a smoothing effect, the priors for the spline coefficients $a_{p}$, \(p>1\), are considered to be normal distributions \(a_{p} \sim \mathcal{N}(a_{p-1},\, \theta)\) with the previous spline coefficients \(a_{p-1}\) as expected values and common variance \(\theta \sim \mathcal{N}^{+}(0,\, 1)\), while the prior for the first spline coefficient $a_{1}\sim \mathcal{N}^{+}(0,\, 1)$ is considered to be a half-normal distribution.
The Bayesian hierarchical model is implemented in R~\cite{rProg} via the add-on package \texttt{rstan} for Stan~\cite{rStan}. The implementation of smoothing splines is based on Kharratzadeh~\cite{kharratzadeh2017splines}. 
Posterior samples are obtained by the No-U-Turn Sampler for Hamiltonian Monte Carlo~\cite{hoffman2014, betancourt2015}, using eight independent chains with 2000 iterations each (considering burn-in periods of 1000 iterations). Convergence diagnostics indicate that the algorithm provides a representative sample from the posterior distribution (see Section~\ref{ssec:diagnostics} of the Supplement).  

\section{Results} \label{sec:results}

 Using German surveillance data~\cite{rkisit2021, rkitests}, we model the course of the pandemic for Germany as a whole country as well as separately for each of the 16 German federal states during the first year of the pandemic. Here we present detailed national results, while further results for the individual federal states can be found in Section~\ref{ssec:states} of the Supplement. 

\begin{figure}[th!]
 \centering
 \includegraphics[width=\textwidth]{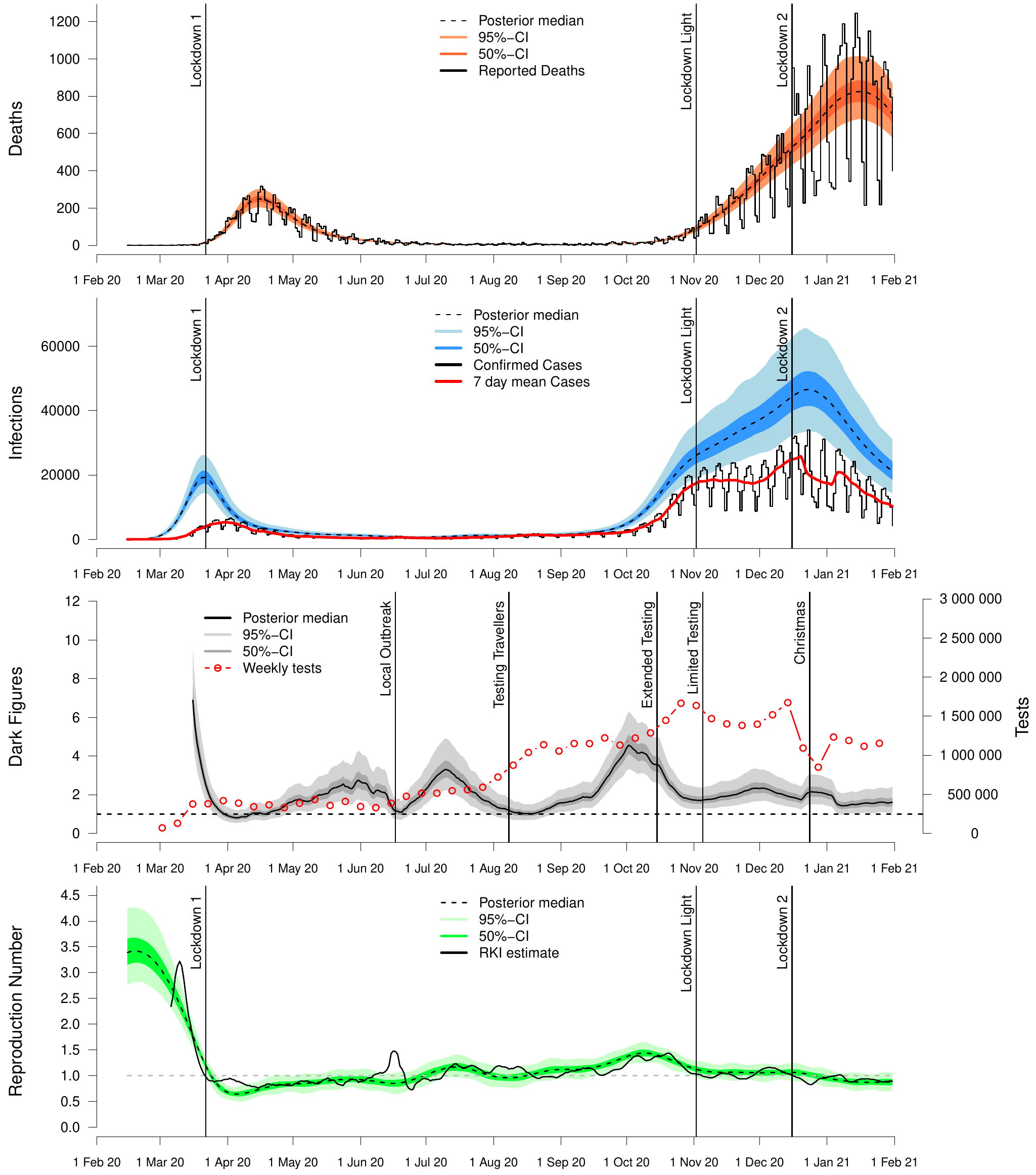}
 \caption{Results for the course of the COVID-19 pandemic in Germany via spline-based hierarchical modelling of death counts (cf.\ Figure~\ref{fig:modelflow}), based on age-specific IFR estimates from Brazeau et al.~\cite{brazeau2020}. Model estimates are based on posterior medians, together with 50\% and 95\% credible intervals (CIs).}
 \label{fig:pdeBR}
\end{figure}

Results of the Bayesian hierarchical model (cf.\ Figure~\ref{fig:modelflow}) for Germany are depicted in Figure~\ref{fig:pdeBR}. During the first year of the pandemic, death counts show two pronounced waves, with a shorter first wave in spring~2020 with fewer deaths compared to the second wave in autumn and winter 2020/2021. Figure~\ref{fig:pdeBR} (first plot) indicates that the hierarchical model yields a good fit to the course of reported numbers of deaths. Note that daily reported deaths and confirmed cases show a characteristic weekly oscillating pattern. Estimates of our model, however, capture the average tendency and are robust to such reporting artefacts due to the use of smoothing splines with biweekly equidistant knots  for modelling the effective reproduction number.

Figure~\ref{fig:pdeBR} additionally illustrates the introduction of major non-pharmaceutical interventions (NPIs) in Germany. Note that, due to the federal structure of Germany, there have been local differences regarding the implementation of NPIs (not shown here). Around the 23rd of March 2020, the ``first lockdown'' (Lockdown~1) was introduced in Germany, including strict contact restrictions and closures of non-essential shops. At this point, schools and nurseries had already been closed for one week and public events had been banned for two weeks in most parts of the country. Approximately three to four weeks after the introduction of the ``first lockdown'', the first wave reached its maximum regarding the number of reported deaths. During summer 2020, cases and death counts were at relatively low levels. After a rapid increase of cases in October 2020, the German government introduced the so-called ``lockdown light'' to mitigate the spread of the virus. During this period, restaurants and leisure facilities were closed, while schools and shops remained opened with hygiene concepts in place. However, no noticeable decline of reported deaths was observed and confirmed cases remained at relatively high levels, which led to the introduction of the ``second lockdown'' in December 2020 (Lockdown~2), including closures of schools and non-essential shops amongst further measures. Three to four weeks after the introduction of the ``second lockdown'', a decline of daily deaths was reported similarly as during the first wave in spring~2020. Note that, as expected, the general course of confirmed cases precedes the course of reported deaths. 

Instead of modelling directly the numbers of confirmed cases, which are largely influenced by the adopted testing regime, the Bayesian hierarchical model provides estimates of the course of true infections. Estimated infections in turn precede the course of confirmed cases due to the incubation period and reporting delays (see second plot of Figure~\ref{fig:pdeBR}). However, results show a remarkable difference between the course of estimated infections and confirmed cases during the second wave of infections in autumn/winter 2020/21: following the ``lockdown light'' introduced at the 2nd of November~2020, the course of confirmed cases suggests a flattening or even decreasing trend, while results of our hierarchical model indicate that true numbers of estimated infections continued to rise and reached a maximum around one week after the ``second lockdown'' in December~2020. 

Such effects should also be viewed in light of changing testing policies, which can result in varying dark figures. In fact, by modelling the true infections, our approach allows for the interpretation of dark figures of infections during the pandemic. During the first year of the pandemic (i.e.\ between the 1st of February~2020 and the 31st of January~2021), there were in total $2,221,838$ confirmed SARS-CoV-2 cases in Germany; for the same time period, the model estimates $4,448,889$~(95\% credible interval: $[3,244,647;\:6,176,642]$) infections based on age-specific IFR estimates from Brazeau et al.~\cite{brazeau2020}, yielding an overall estimated dark figure factor of $2.002\:[1.460;\:2.780]$. Using alternative age-specific IFR assumptions based on O'Driscoll et al.~\cite{o2021} and Levin et al.~\cite{levin2020} (cf. Figure~\ref{fig:ifr-curves} of the Supplement), the model estimates $5,577,165\: [3,965,708;\:7,930,536]$ and $3,008,303\: [2,154,854;\:4,274,711]$ infections for the first year of the pandemic, resulting in overall estimated dark figure factors of $2.510\:[1.785;\:3.569]$ and $1.354\:[0.970;\:1.924]$, respectively. Although absolute numbers of estimated infections and levels of dark figures vary considerably for the different age-specific IFRs, general temporal trends of estimated infections and dark figures remain quite stable, while effective reproduction numbers estimated by our model are very robust regarding different age-specific IFR assumptions (see Section~\ref{ssec:sensitivity} of the Supplement for detailed results of sensitivity analyses).    

The third plot of Figure~\ref{fig:pdeBR} depicts the estimated course of the dark figures factor, which is given by the ratio of numbers of estimated infections divided by numbers of confirmed cases. Since there is an average delay of ten days between infections and reporting dates (see Methods), dark figures for day~$t$ are computed as the ratio of estimated infections at day~$t-10$ and the 7-day-mean of detected cases at day~$t$. Results show that estimated dark figures are much larger before the ``first lockdown'' in spring~2020 than in the following course, reflecting limited testing capacities at the beginning of the pandemic. It can be observed that changes in estimated dark figures are often associated with shifts in testing policies and practice (here, vertical lines in the third plot of Figure~\ref{fig:pdeBR} indicate dates of important shifts). In particular, estimated dark figures declined sharply in June~2020 in the context of a local super-spreading event in a slaughterhouse in North Rhine-Westphalia, resulting in temporarily increased targeted testing of factory employees. During August~2020, when incidence rates were low and targeted testing of travellers returning from summer holidays was intensified, the factor of dark figures is estimated to be close to one. Following increasing dark figures of infections in September and October~2020, on the 15th of October the Robert Koch Institute~(RKI) changed its recommendations towards less stringent indications for SARS-CoV-2 tests, leading to a further increase in numbers of conducted tests and a considerable decrease in estimated dark figures of infections. However, with increasing incidences towards the end of October, on the 5th of November the recommendations were again updated towards more restricted testing, which again results in increasing estimated dark figures from our model. Finally, temporarily increased dark figures are estimated following Christmas, reflecting the decrease of conducted tests during this holiday period. Overall, our results indicate that the hierarchical modelling approach can reliably detect shifts in testing policies, even though model estimates of dark figures are solely based on numbers of confirmed cases and reported deaths, without incorporating any direct information regarding the numbers of conducted tests.

Finally, the last plot of Figure~\ref{fig:pdeBR} shows the resulting course of the effective reproduction number estimated by our hierarchical model based on death counts, in comparison with official estimates from the Robert Koch Institute~(RKI), which are based on the evolution of confirmed cases~\cite{heiden2020, rkiRwert2021}. It can be observed that estimates based on death counts are often similar to classical estimates based on confirmed cases. However, model estimates based on death counts prove to be more robust against shifts in testing policies. In particular, confirmed cases indicate a short-term spike in the effective reproduction number linked to a local super-spreading event in June~2020, whereas our model does not estimate a spike during this period; instead, it estimates reduced dark figures of infections, suggesting that the spike in the effective reproduction number was mainly related to targeted testing of contact persons. 
Furthermore, after the ``lockdown light'' at the 2nd of November~2020, classical estimates of the effective reproduction number tend to be smaller than model estimates based on death counts. Although the differences do not seem large, they imply considerably different interpretations regarding the course of the pandemic: while classical estimates (with values smaller or equal to one) suggest a flattening or even decreasing trend of infections following the ``lockdown light'', estimates of our model (with values larger or equal to one) suggest that true numbers of infections continued to rise (cf.\ second plot of Figure~\ref{fig:pdeBR}). Note that the ``lockdown light'' was introduced more or less at the same time when the RKI recommendations for testing were adapted (see third plot of Figure~\ref{fig:pdeBR}), indicating that our model is able to disentangle overlying effects of reduced testing (resulting in larger dark figures) and the adaptation of NPIs on numbers of confirmed cases.

\section{Discussion} \label{sec:discussion}
  
We have extended and adapted the Bayesian hierarchical model of Flaxman et al.~\cite{flaxman2020estimating} for modelling the course of the COVID-19 pandemic in Germany based on death counts. 
A main feature of the proposed approach is the smooth and data-driven way of estimating the effective reproduction number. As a result, there is no need to prespecify discrete change points for timings of non-pharmaceutical interventions (NPIs) as in the original model of Flaxman et al.~\cite{flaxman2020estimating}, diminishing the chance that potential implicit biases are incorporated into the model~\cite{chin2021, wood2021}. While our approach shows parallels with the Bayesian model developed in Wood~\cite{wood2021}, which uses smoothing splines to estimate effective reproduction numbers in the United Kingdom, an important additional characteristic of our work is the adjustment for time-varying effective infection fatality rates (IFRs), which are estimated to change substantially over the course of the pandemic as a result of varying age distributions of infections (cf.~\cite{staerk2021}).  

Results for German surveillance data illustrate that the proposed retrospective modelling  can provide additional valuable insights regarding the course of effective reproduction numbers and dark figures of true infections. While estimated reproduction numbers of our model are often similar to classical estimates from the Robert Koch Institute~(RKI) based on confirmed cases~\cite{heiden2020, rkiRwert2021}, the proposed modelling approach based on death counts proves to be more robust against shifts in testing policies. In contrast to classical estimation methods relying solely on confirmed cases, our approach has the potential to disentangle overlying effects of shifts in testing policies and actual changes in the effective reproduction number, as illustrated for the second wave of infections in Germany in November~2020, where the ``lockdown light'' was introduced concurrently with the adaptation of testing recommendations. Further results presented in the Supplement illustrate that our Bayesian modelling approach yields robust results for the different developments of the pandemic in the 16 individual German federal states (using the same model and priors as for the full country). Here one should note that death counts were at relatively low levels during summer~2020, so that estimating the effective reproduction number and dark figures of infections for individual federal states comes with larger uncertainty, which is also reflected in the wider credible intervals of model estimates (particularly for federal states with smaller populations, see e.g.\ the results for Mecklenburg-Western Pomerania and Saarland in Figures~\ref{fig:Mecklenburg} and~\ref{fig:Saarland}). 

Our study is also related to the recent work of Schneble et al.~\cite{schneble2021}, which estimates relative changes in the case detection ratio~(CDR) over time for different age groups. The authors employ a smooth generalized linear mixed model for confirmed cases and variables indicating whether the infections resulted in COVID-19 related deaths (without modelling the actual dates of deaths). While the classical mixed modelling approach provides age-group specific estimates of \textit{relative} changes in the CDR, our proposed Bayesian hierarchical modelling approach also yields estimates of \textit{absolute} numbers of dark figures of infections as well as estimates of the effective reproduction number, by considering age-specific IFR estimates and dates of reported deaths. Our numerical results regarding trends in dark figures of infections generally support the results of Schneble et al.~\cite{schneble2021}: dark figures in Germany are estimated to be largest at the beginning of the pandemic and, after a period of relatively low estimated dark figures (i.e.\ large CDR) during summer~2020, numbers of undetected cases are estimated to  increase sharply in September~2020.

An obvious limitation of our modelling approach relying on death counts is that it can only reflect the course of the pandemic in retrospect. While this is partly true for all modelling approaches, including the traditional ones based on confirmed cases (due to the incubation period and reporting delays), one clearly has to acknowledge that a timely reporting and analysis is essential for estimating the effects of political decisions~(e.g., lockdown measures or other NPIs). In this context, models based on death counts will always yield results several weeks later than day-by-day estimates relying on reported cases. 
A limitation of the presented analysis of German surveillance data is that dates of daily reported deaths may deviate from actual dates of deaths. Another limitation of the Bayesian hierarchical modelling approach is that it relies on various assumptions (cf.~\cite{flaxman2020estimating}), among them statistical ones including the implemented prior distributions for model parameters (see Section~\ref{sec:methods}). From a more practical perspective, the model also relies on the assumption of a closed environment (no new infections imported from outside of the population) and on the assumption that cases are insusceptible for another (second) infection with COVID-19. For the estimation of the effective infection fatality rate~(IFR) we assumed that the evolving age distribution of infections can be approximated by the corresponding age distribution of confirmed cases (cf.~\cite{staerk2021}); furthermore, it relies on the assumption that age-specific IFR estimates from Brazeau et al.~\cite{brazeau2020} are applicable to Germany (see Section~\ref{ssec:sensitivity} of the Supplement for sensitivity analyses with alternative IFR estimates). In the current modelling approach we do not account for vaccinations, which does not pose an important limitation for the considered time period with low total numbers of administered vaccinations until the end of January~2021.

In this context, we have reason to believe that our hierarchical approach is particularly flexible regarding extensions for future challenges of COVID-19 modelling, as it shifts the focus from modelling confirmed cases towards reported deaths. Since we have included time-varying IFRs to account for changing age distributions of infections, future research could make use of this flexibility and incorporate also time-varying vaccination effects into our model. This could be achieved via a general population-wise factor or age-group specific parameters representing the rates of fully vaccinated individuals in the respective groups. Especially in light of progressive vaccination programs in many countries, it can be expected that there will be additional sharp changes in implemented testing regimes during the further course of the pandemic; for example, at the time of writing it was announced that SARS-CoV-2 tests will generally be provided for free in Germany only until the 11th of October~2021. In addition, vaccinated individuals may generally be less likely to be tested due to asymptomatic or mild symptomatic infection (cf.~\cite{rovida2021}), which may induce larger dark figures of infections in the vaccinated part of the population. Such vaccination effects are expected to further complicate the reliable estimation of the effective reproduction number based only on the numbers of confirmed cases. Therefore, future research should be targeted at incorporating additional information for modelling the further course of the pandemic, including data on vaccinations, hospitalizations and death counts.

\bibliographystyle{vancouver}
\bibliography{Literatur}

\clearpage

\appendix
\beginsupplement

\section*{Supplement} \label{sec:supplement}

\subsection{Effective infection fatality rate~(IFR)} \label{ssec:effective-ifr}

\begin{figure}[h]
 \centering
 \includegraphics[width=\textwidth]{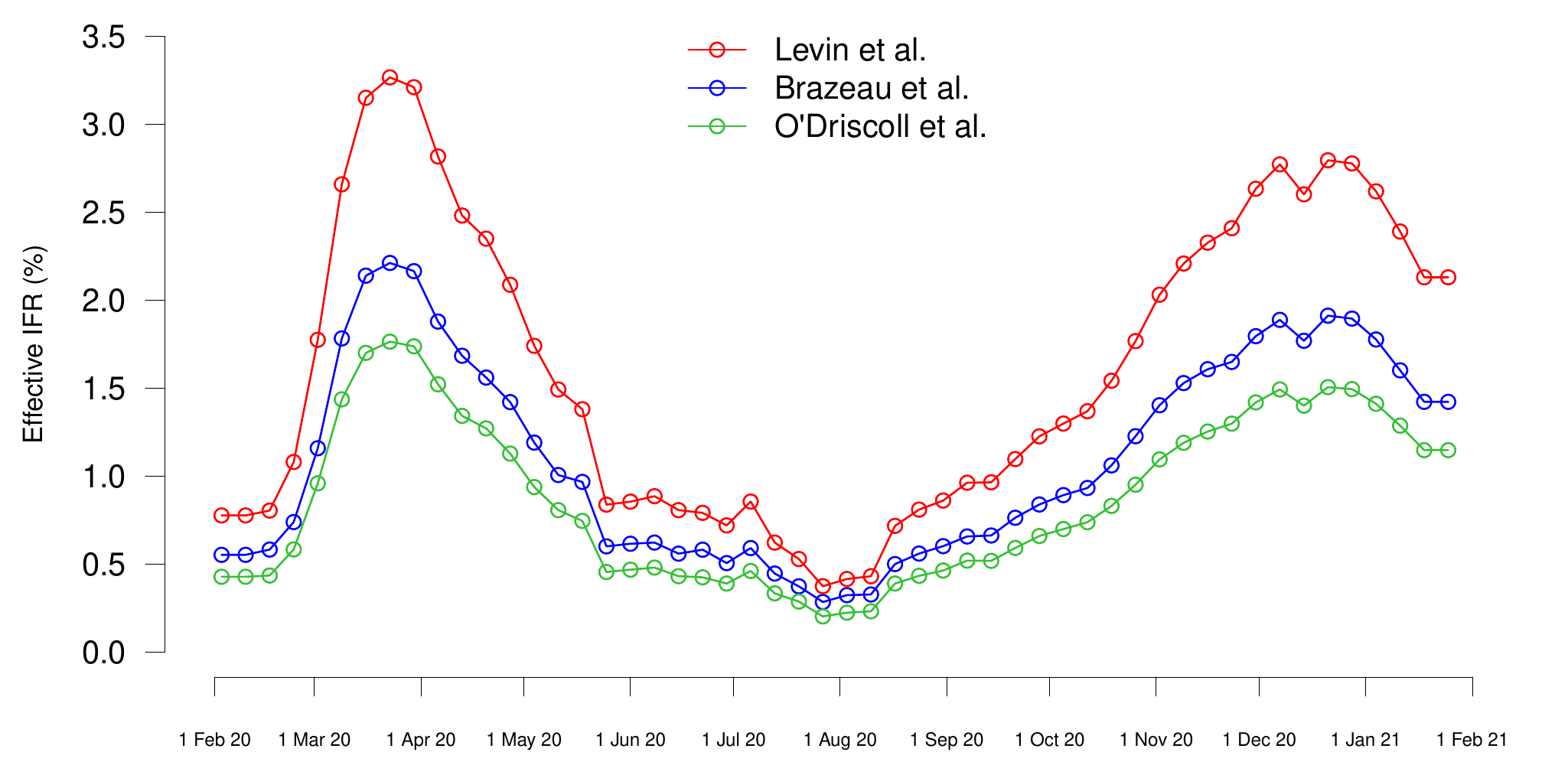}
 \caption{Effective infection fatality rate~(IFR) estimates per calendar week for Germany, based on the three considered international studies providing age-specific IFR estimates (O'Driscoll et al.~\cite{o2021}, Levin et al.~\cite{levin2020}, Brazeau et al.~\cite{brazeau2020}).} 
 \label{fig:ifr-curves}
\end{figure}

\clearpage

\subsection{Sensitivity analyses with alternative age-specific IFR estimates} \label{ssec:sensitivity}

\begin{figure}[th!]
 \centering
 \includegraphics[width=\textwidth]{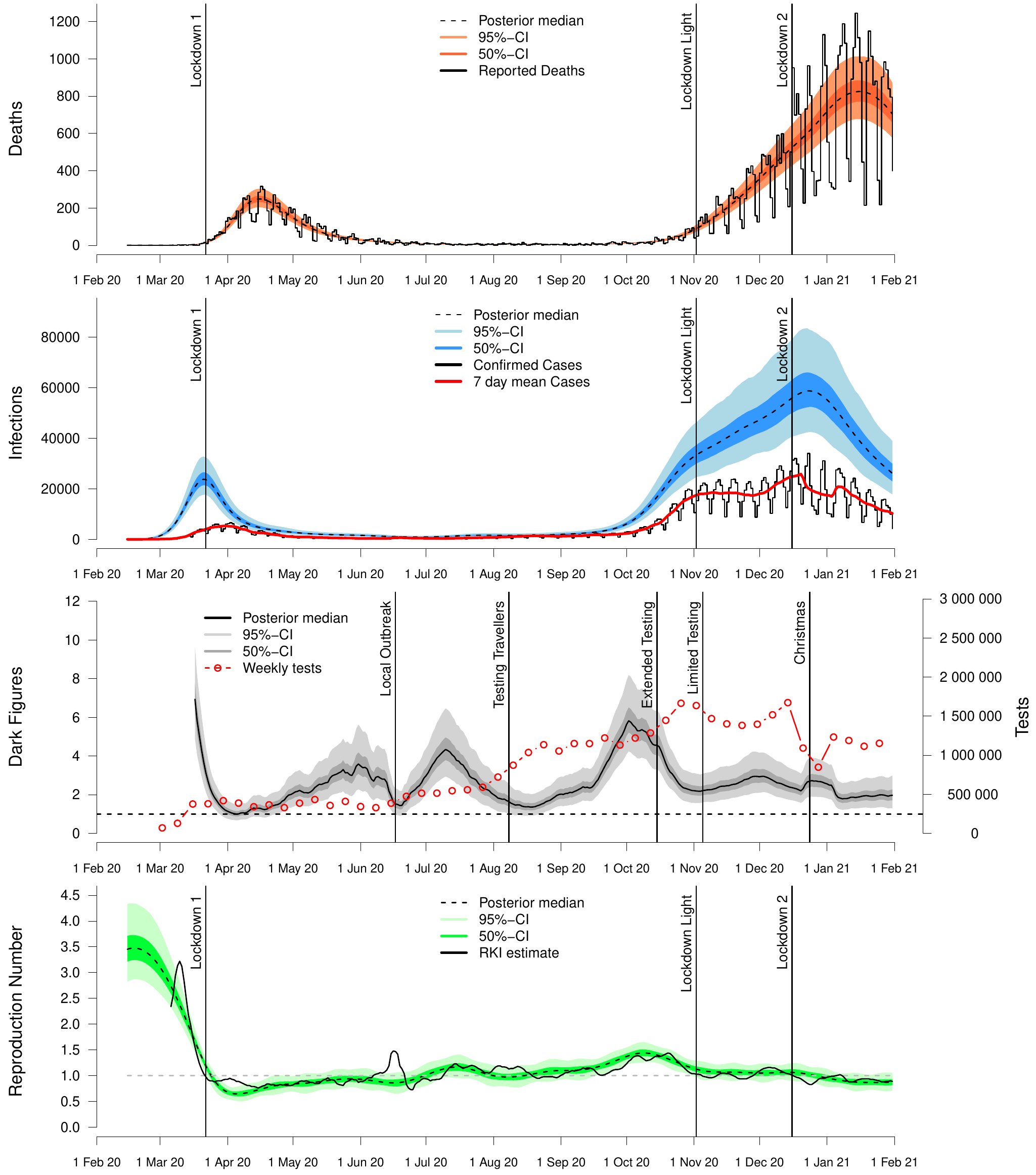}
 \caption{Results for Germany based on age-specific IFR estimates from O'Driscoll et al.~\cite{o2021}.}
 \label{fig:pdeOD}
\end{figure}

\begin{figure}[th!]
 \centering
 \includegraphics[width=\textwidth]{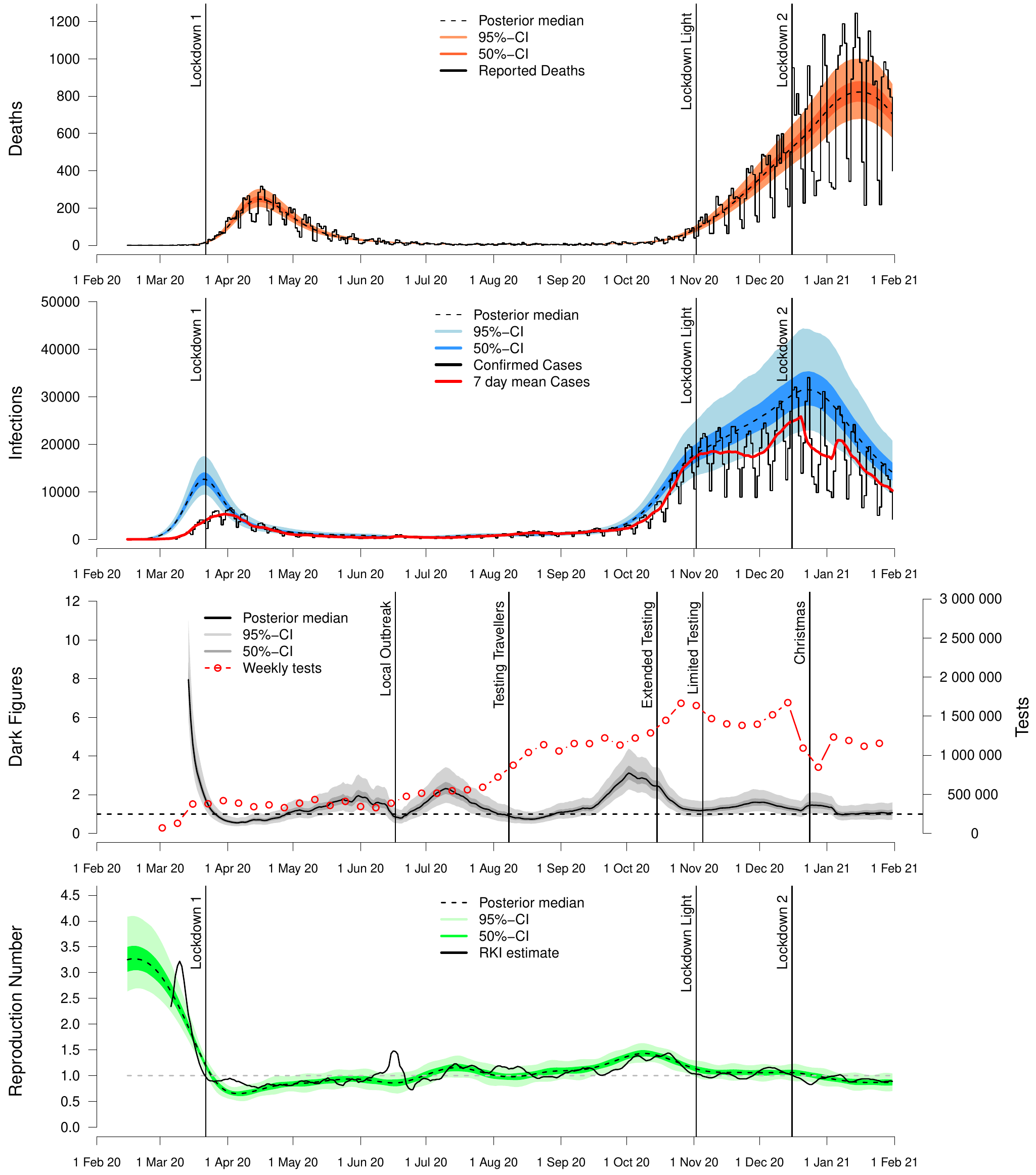}
 \caption{Results for Germany based on age-specific IFR estimates from Levin et al.~\cite{levin2020}.}
 \label{fig:pdeLE}
\end{figure}

\clearpage

\subsection{Results for the 16 German federal states} \label{ssec:states}

\subsubsection{Baden-Württemberg}
\begin{figure}[h]
 \centering
 \includegraphics[width=\textwidth]{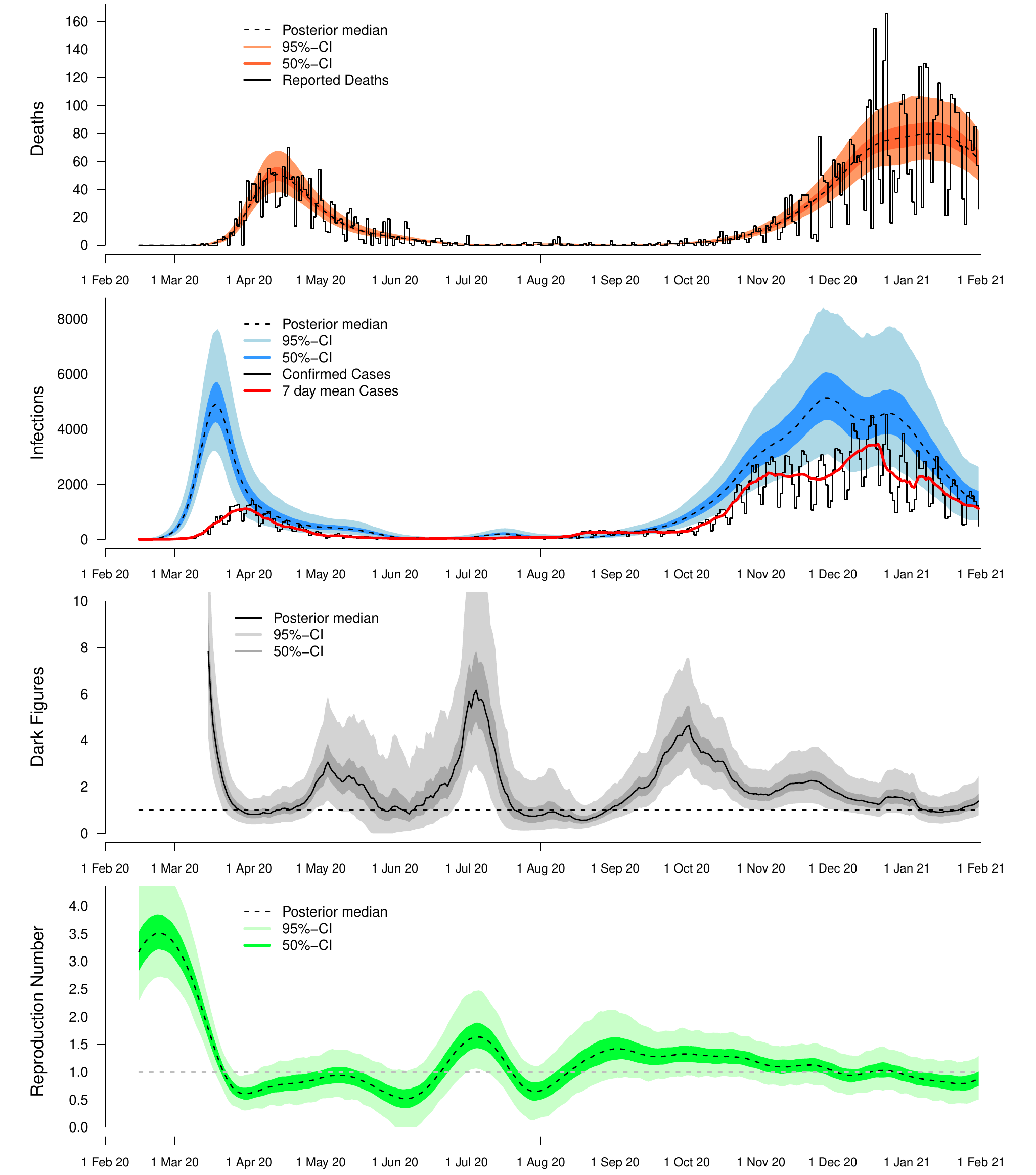}
 \caption{Results for Baden-Württemberg based on age-specific IFR estimates from Brazeau et al.~\cite{brazeau2020}.}
\end{figure}

\clearpage

\subsubsection{Bavaria}
\begin{figure}[h]
 \centering
 \includegraphics[width=\textwidth]{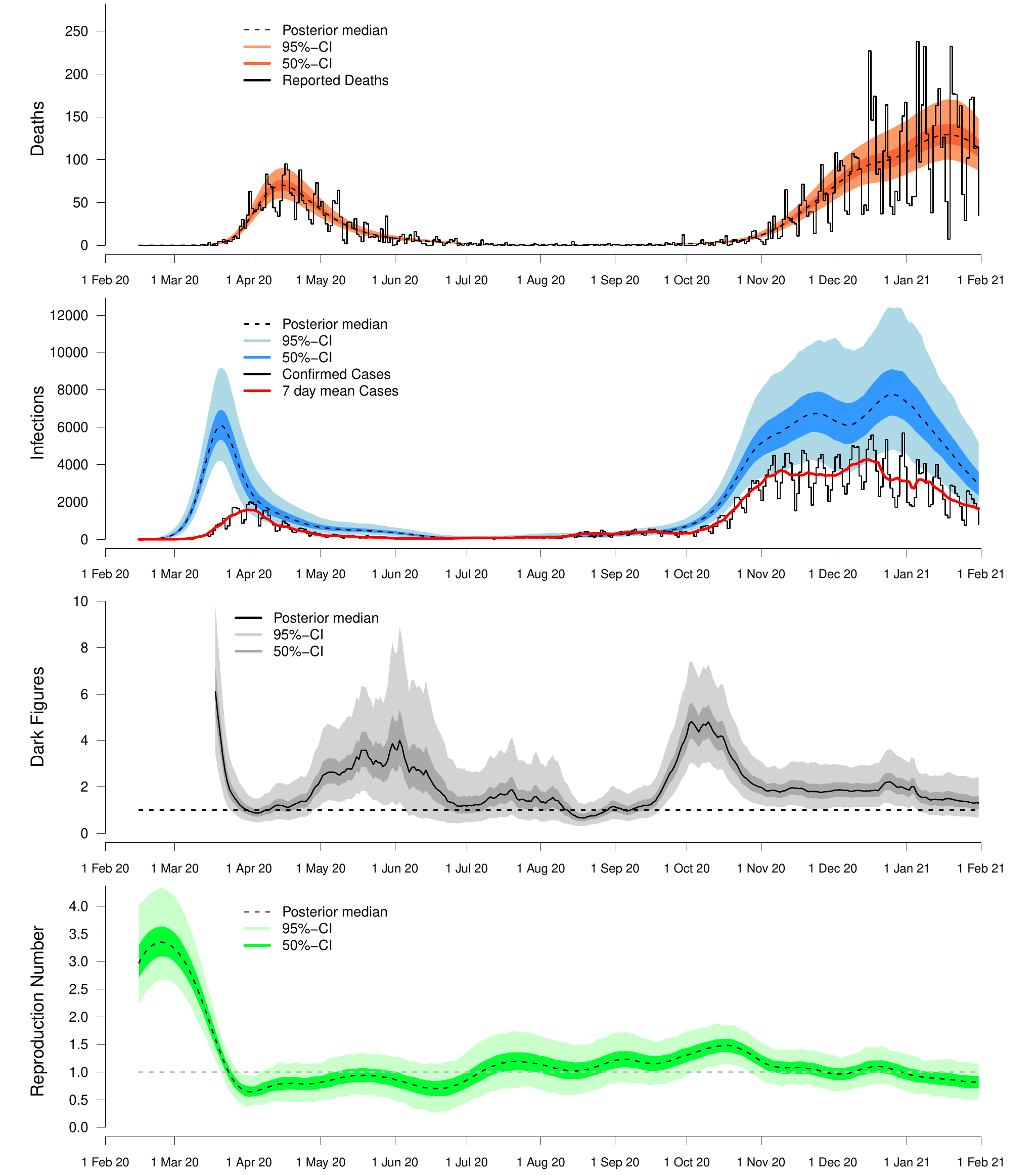}
 \caption{Results for Bavaria based on age-specific IFR estimates from Brazeau et al.~\cite{brazeau2020}.}
\end{figure}

\clearpage

\subsubsection{Berlin}

\begin{figure}[h]
 \centering
 \includegraphics[width=\textwidth]{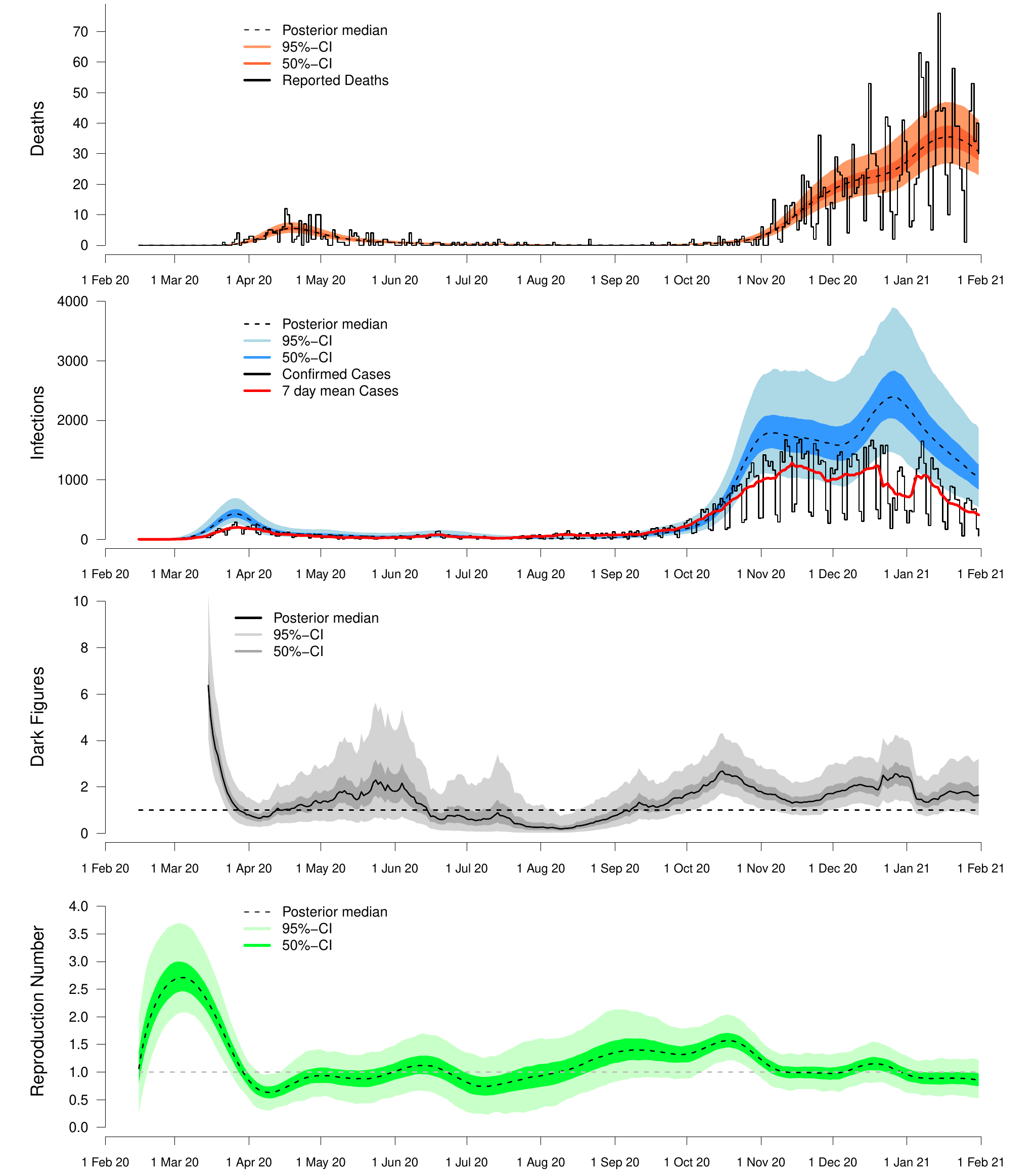}
 \caption{Results for Berlin based on age-specific IFR estimates from Brazeau et al.~\cite{brazeau2020}.}
\end{figure}

\clearpage

\subsubsection{Brandenburg}

\begin{figure}[h]
 \centering
 \includegraphics[width=\textwidth]{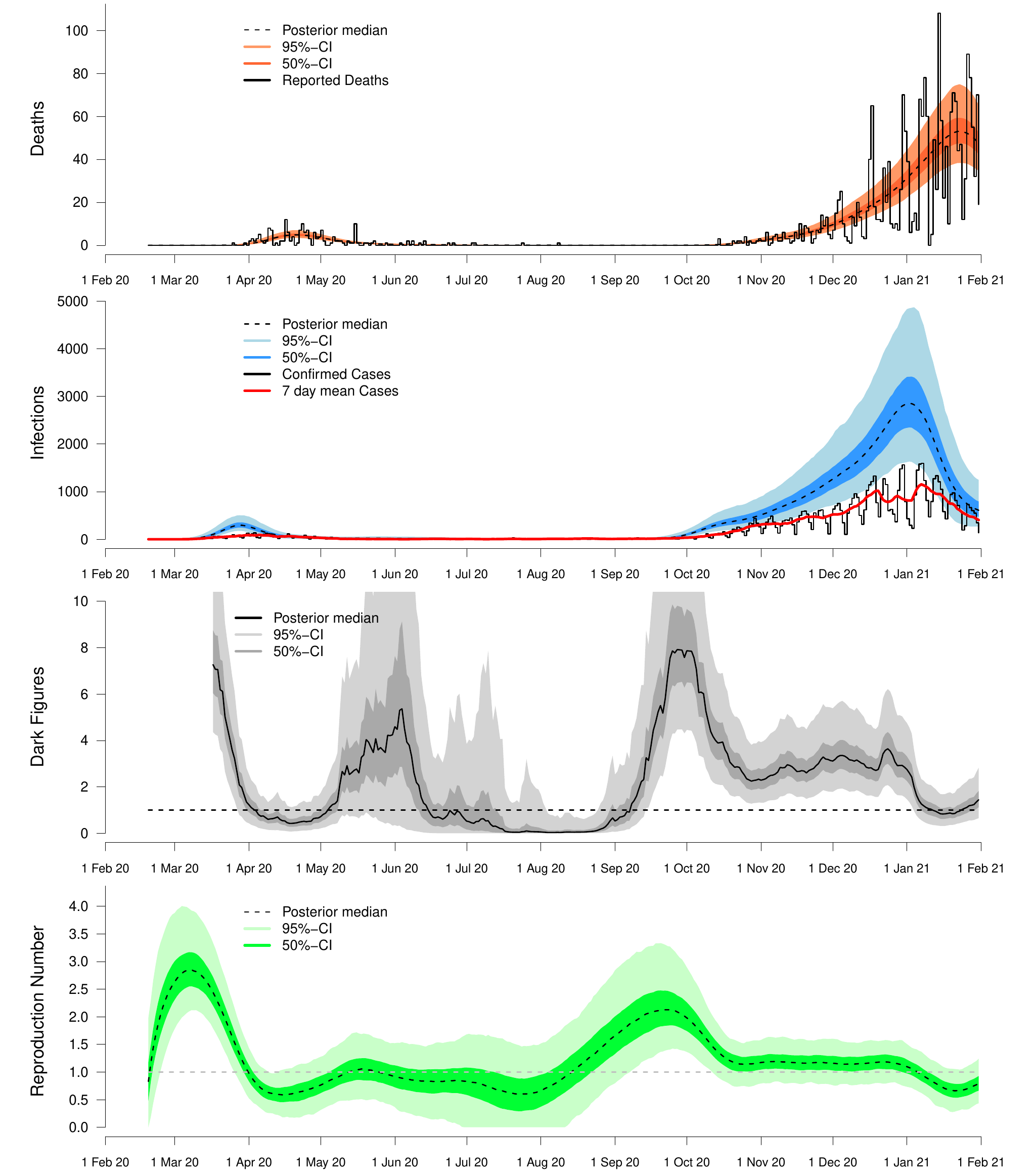}
 \caption{Results for Brandenburg based on age-specific IFR estimates from Brazeau et al.~\cite{brazeau2020}.}
\end{figure}

\clearpage

\subsubsection{Bremen}

\begin{figure}[h]
 \centering
 \includegraphics[width=\textwidth]{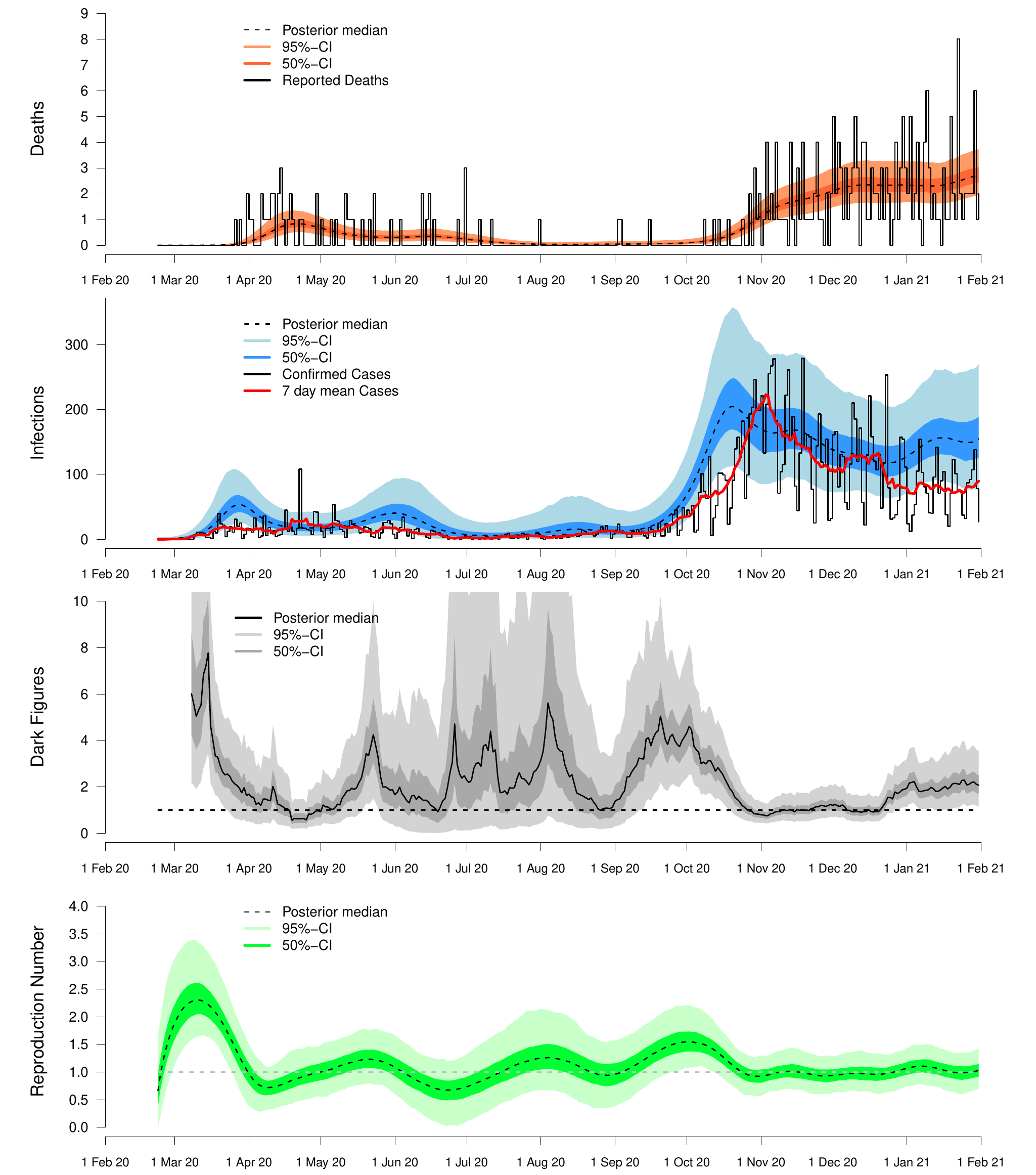}
 \caption{Results for Bremen based on age-specific IFR estimates from Brazeau et al.~\cite{brazeau2020}.}
\end{figure}

\clearpage

\subsubsection{Hamburg}

\begin{figure}[h]
 \centering
 \includegraphics[width=\textwidth]{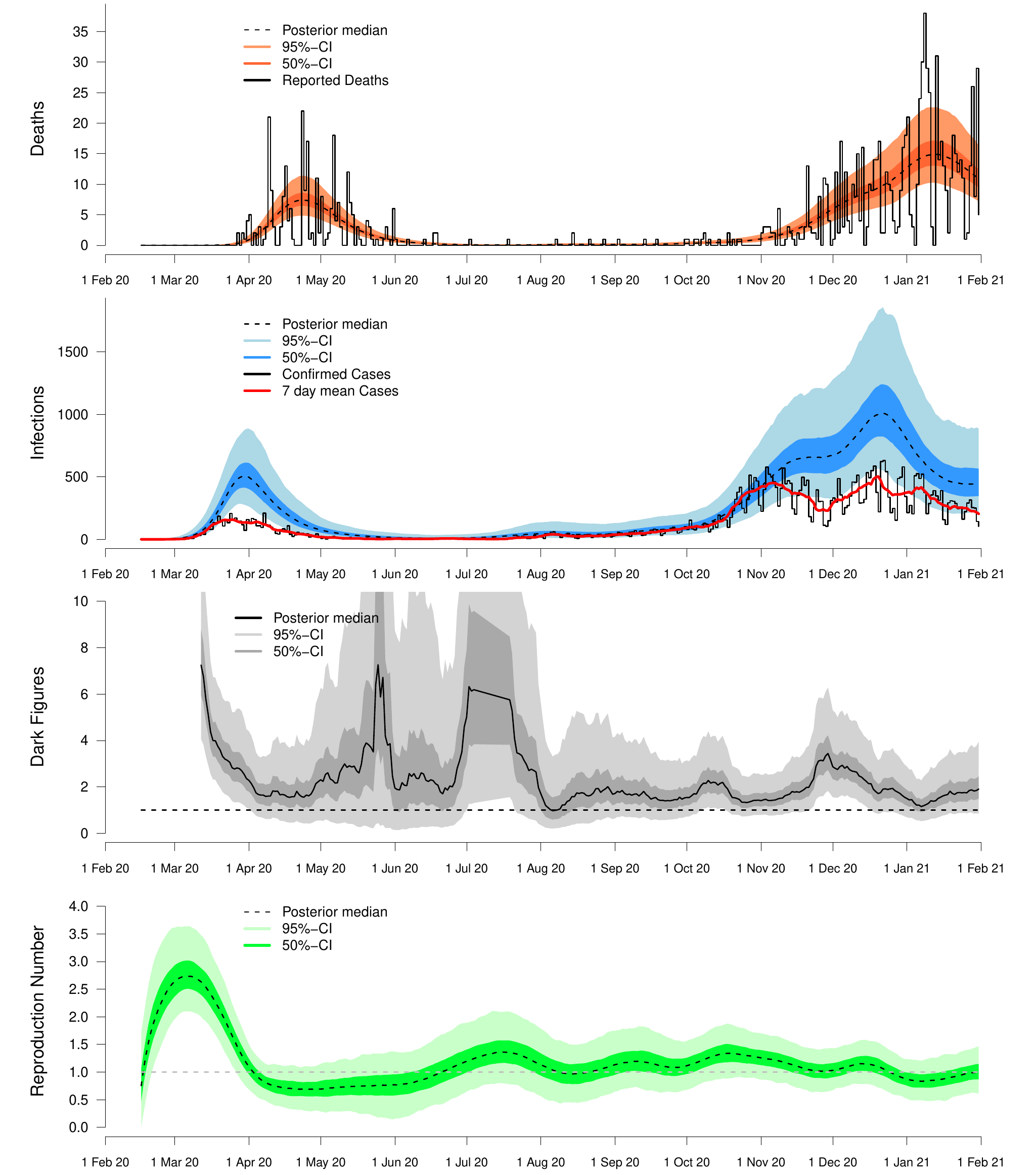}
 \caption{Results for Hamburg based on age-specific IFR estimates from Brazeau et al.~\cite{brazeau2020}.}
\end{figure}

\clearpage

\subsubsection{Hesse}

\begin{figure}[h]
 \centering
 \includegraphics[width=\textwidth]{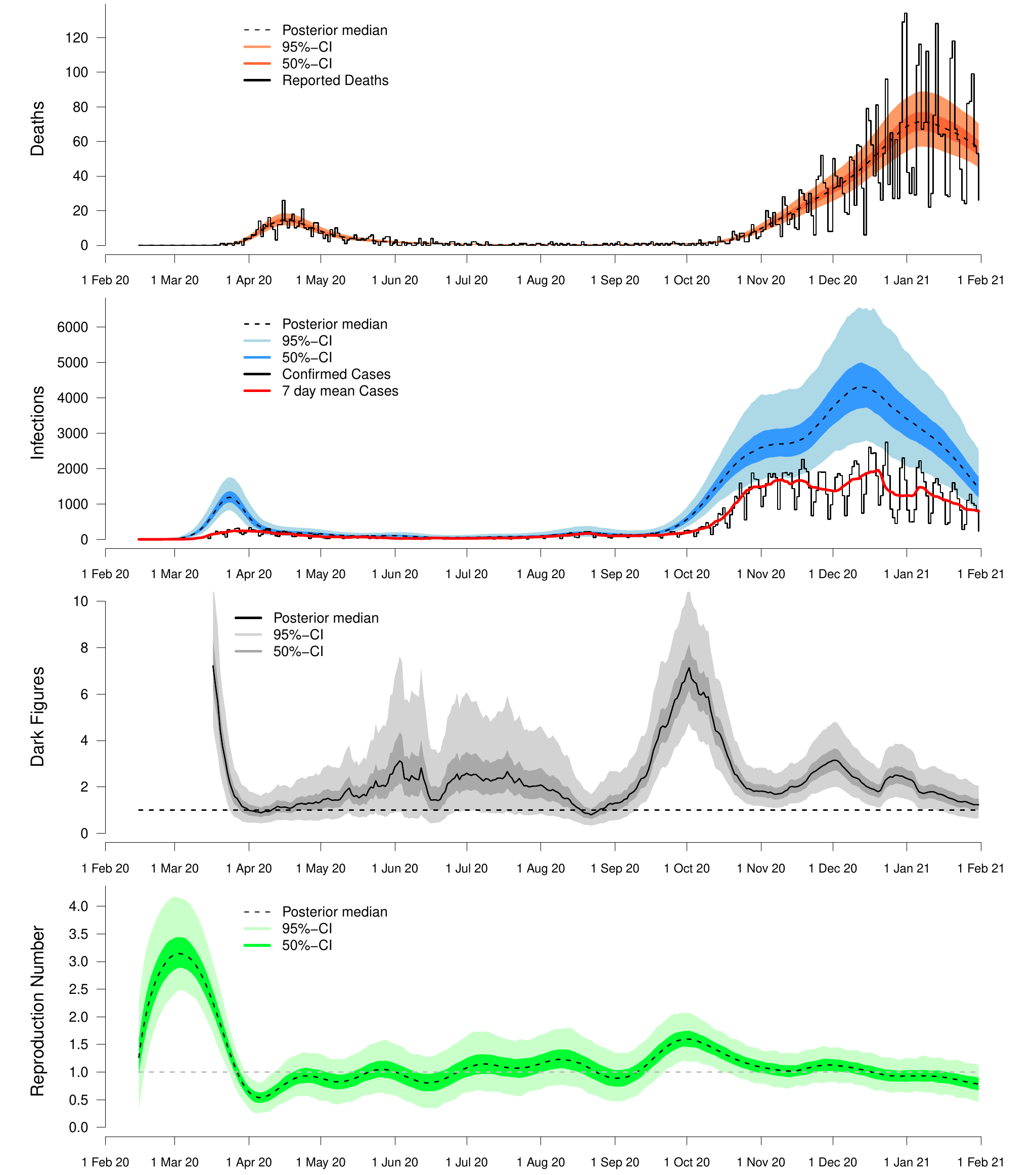}
 \caption{Results for Hesse based on age-specific IFR estimates from Brazeau et al.~\cite{brazeau2020}.}
\end{figure}

\clearpage

\subsubsection{Mecklenburg-Western Pomerania}

\begin{figure}[h]
 \centering
 \includegraphics[width=\textwidth]{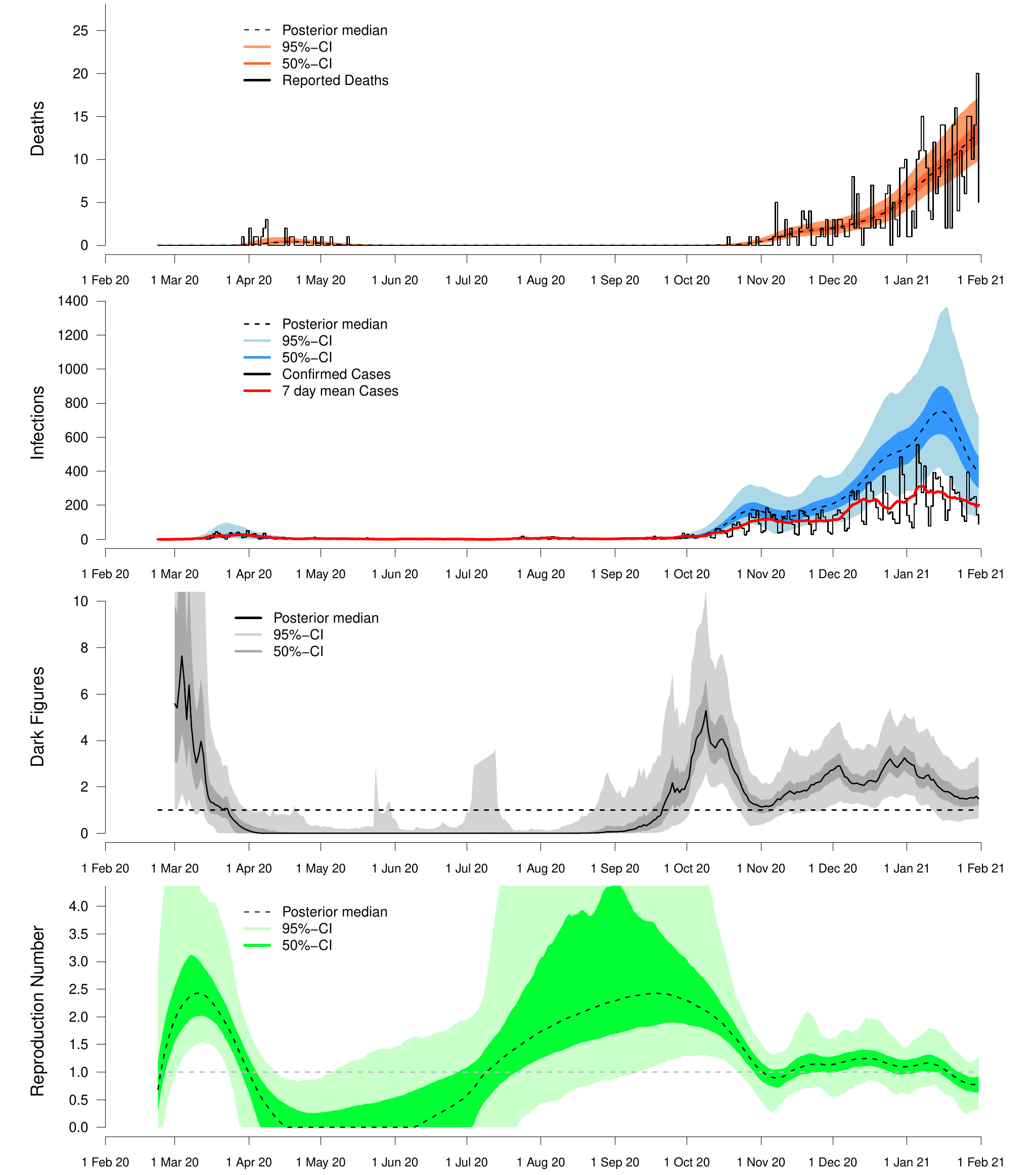}
 \caption{Results for Mecklenburg-Western Pomerania based on age-specific IFR estimates from Brazeau et al.~\cite{brazeau2020}.}
  \label{fig:Mecklenburg}
\end{figure}

\clearpage

\subsubsection{Lower Saxony}

\begin{figure}[h]
 \centering
 \includegraphics[width=\textwidth]{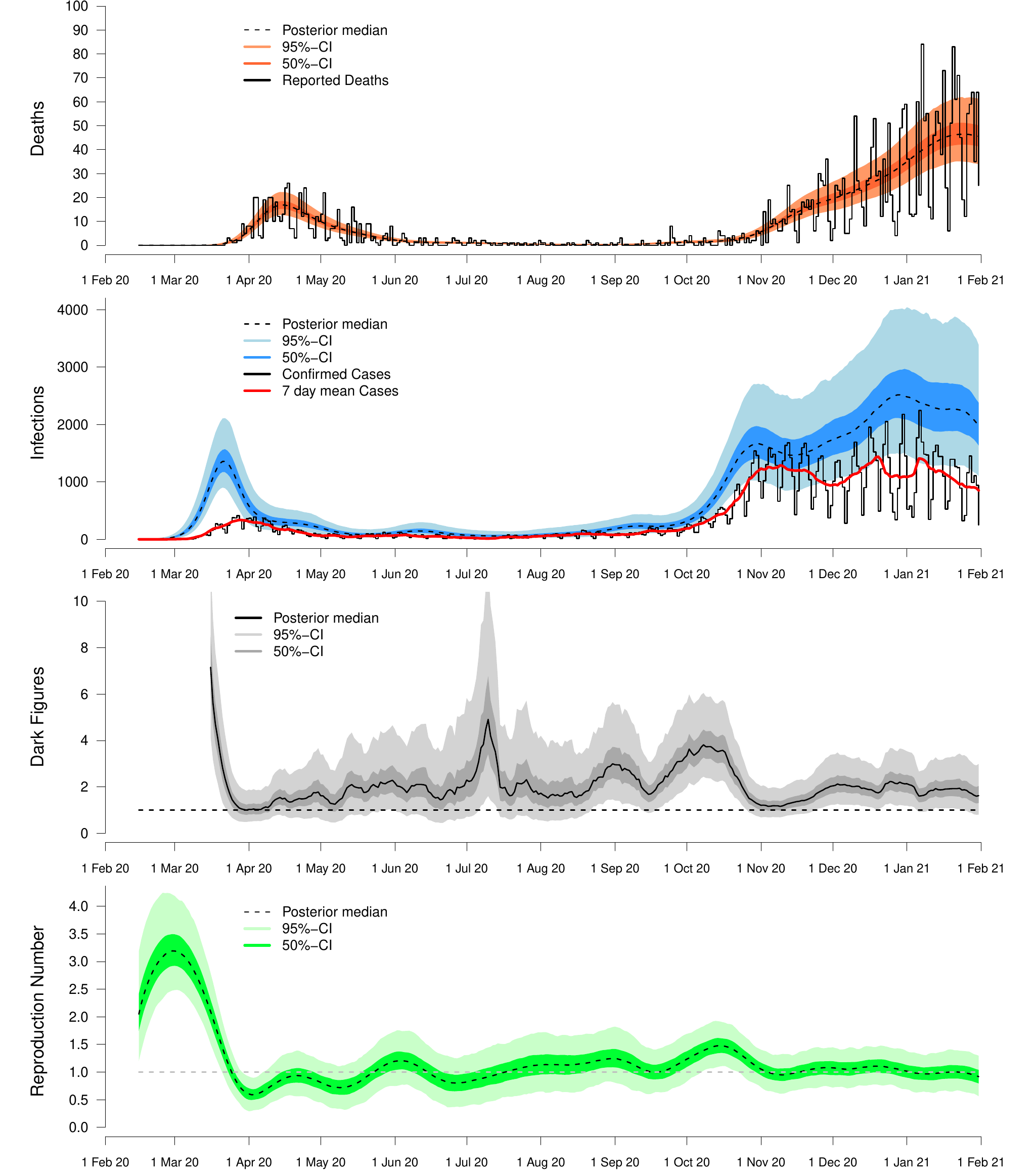}
 \caption{Results for Lower Saxony based on age-specific IFR estimates from Brazeau et al.~\cite{brazeau2020}.}
\end{figure}

\clearpage

\subsubsection{North Rhine-Westphalia}

\begin{figure}[h]
 \centering
 \includegraphics[width=\textwidth]{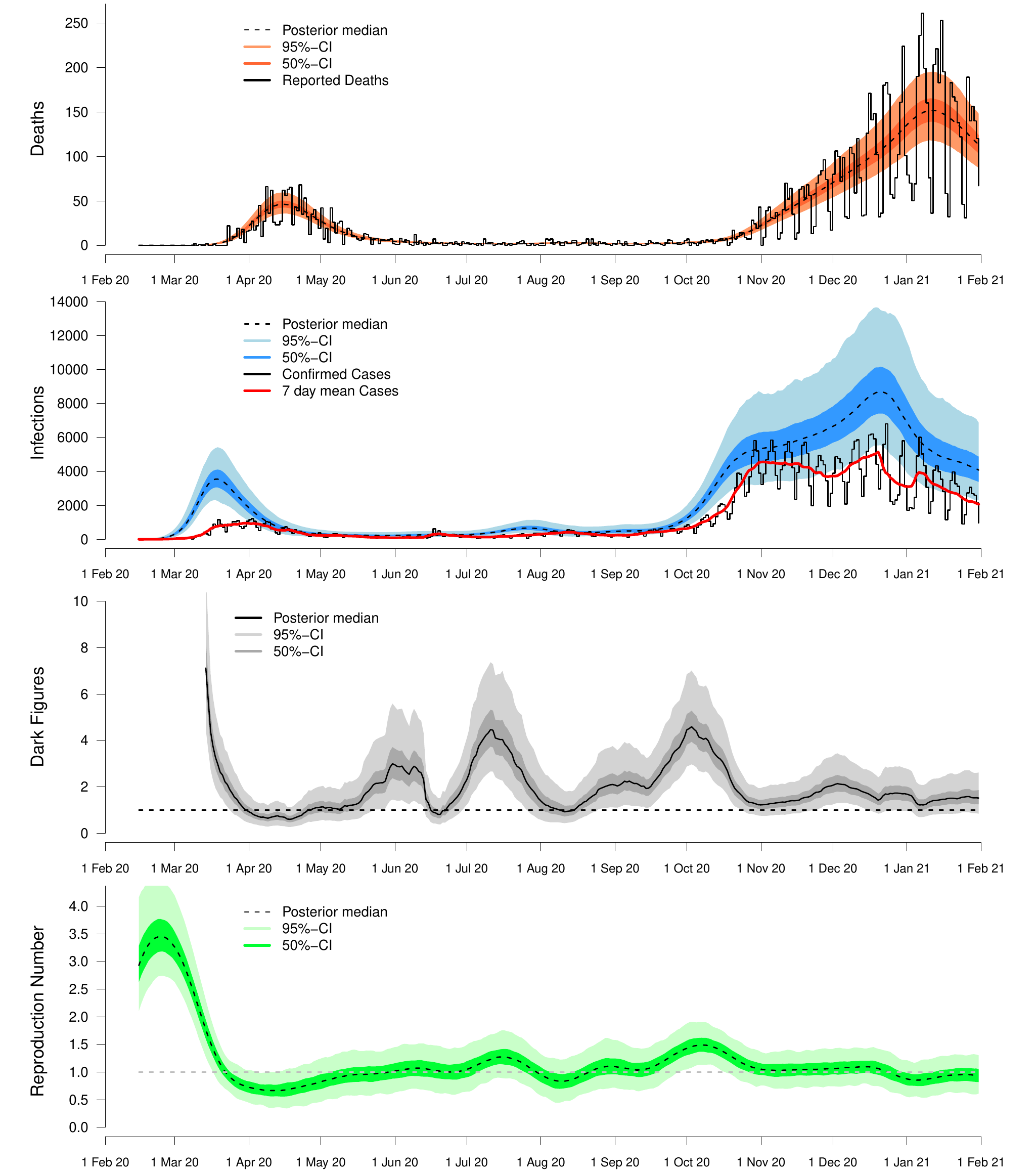}
 \caption{Results for North Rhine-Westphalia based on age-specific IFR estimates from Brazeau et al.~\cite{brazeau2020}.}
\end{figure}

\clearpage

\subsubsection{Rhineland-Palatinate}

\begin{figure}[h]
 \centering
 \includegraphics[width=\textwidth]{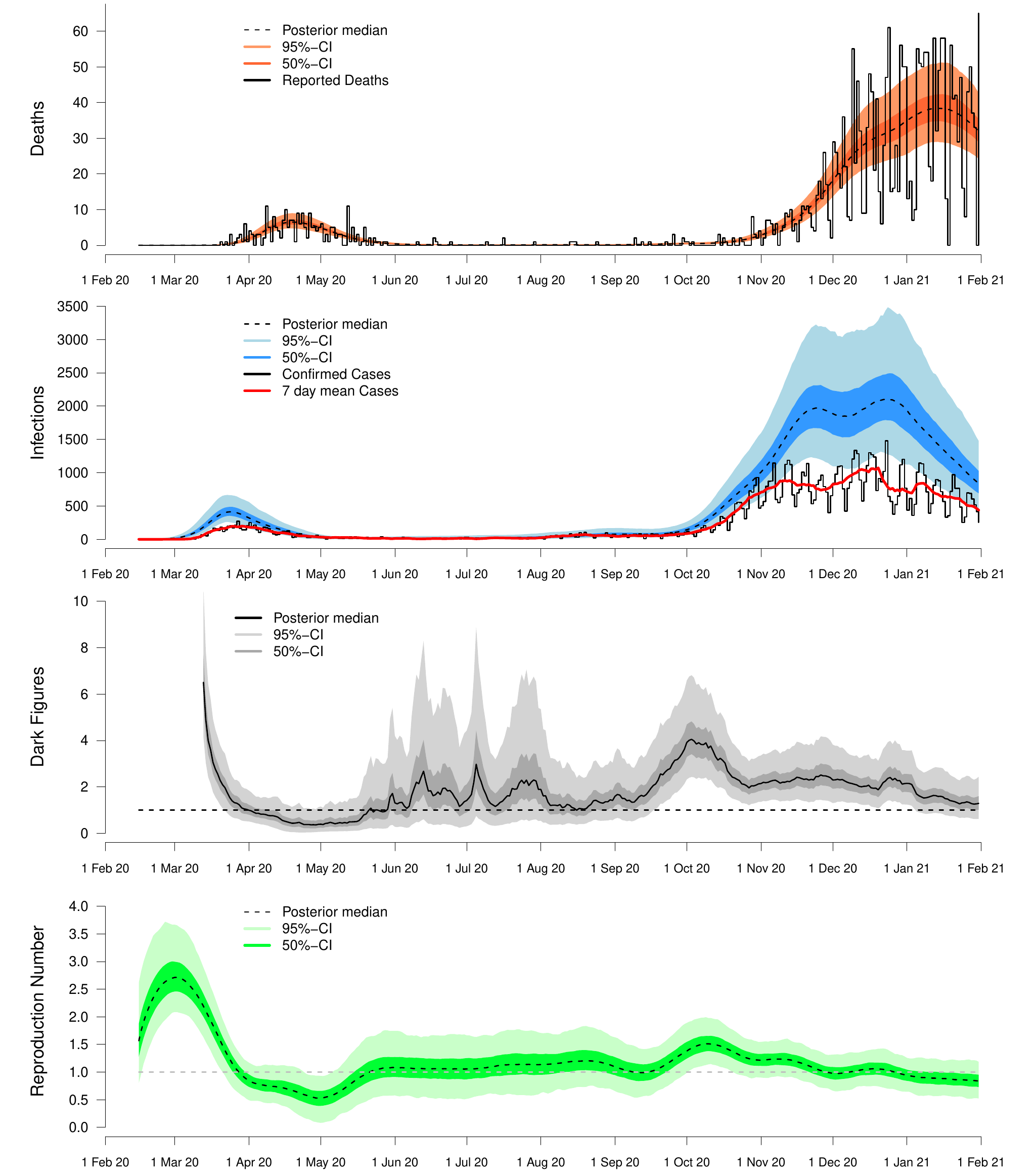}
 \caption{Results for Rhineland-Palatinate based on age-specific IFR estimates from Brazeau et al.~\cite{brazeau2020}.}
\end{figure}

\clearpage

\subsubsection{Saarland}

\begin{figure}[h]
 \centering
 \includegraphics[width=\textwidth]{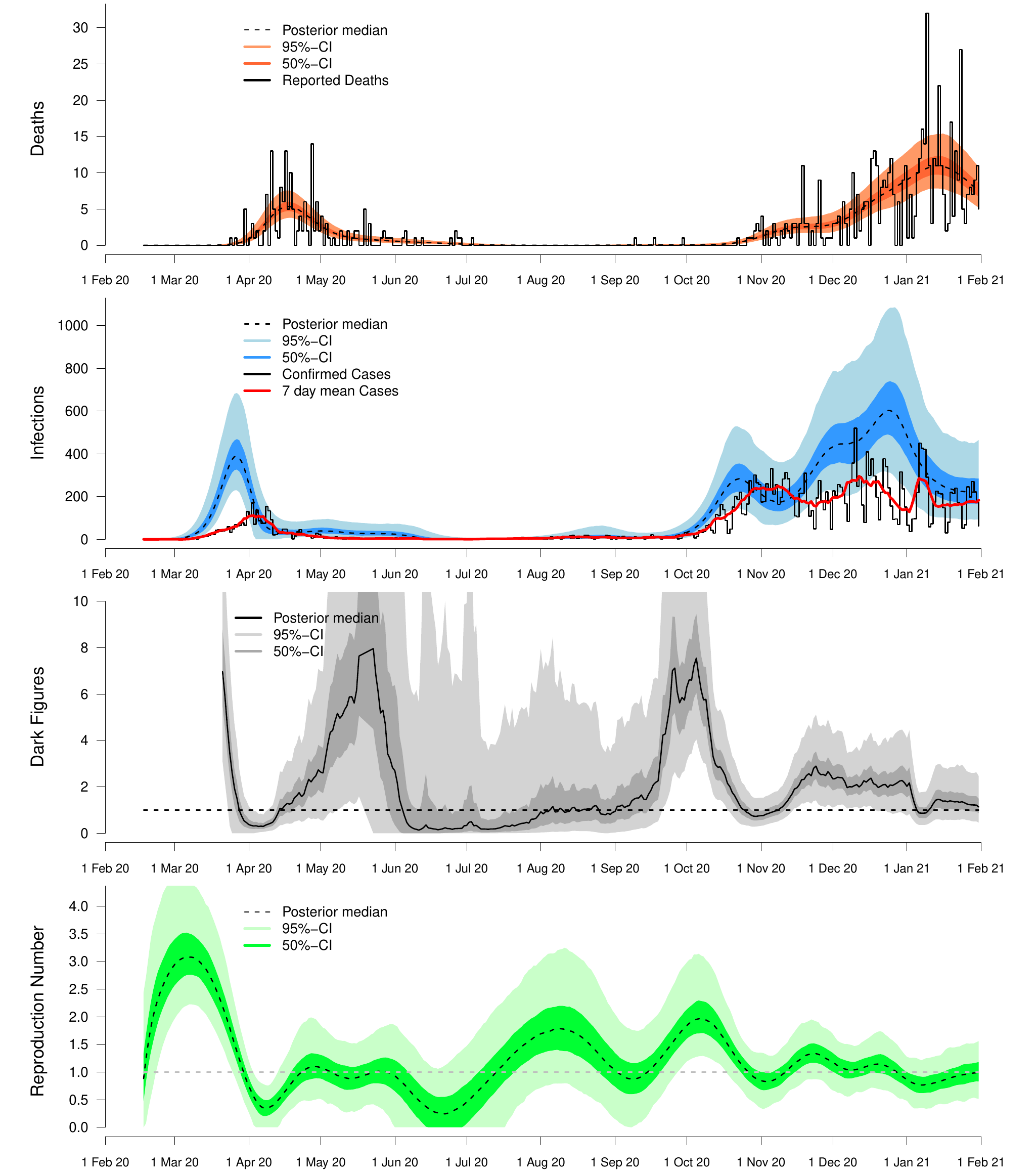}
 \caption{Results for Saarland based on age-specific IFR estimates from Brazeau et al.~\cite{brazeau2020}.}
  \label{fig:Saarland}
\end{figure}

\clearpage

\subsubsection{Saxony}

\begin{figure}[h]
 \centering
 \includegraphics[width=\textwidth]{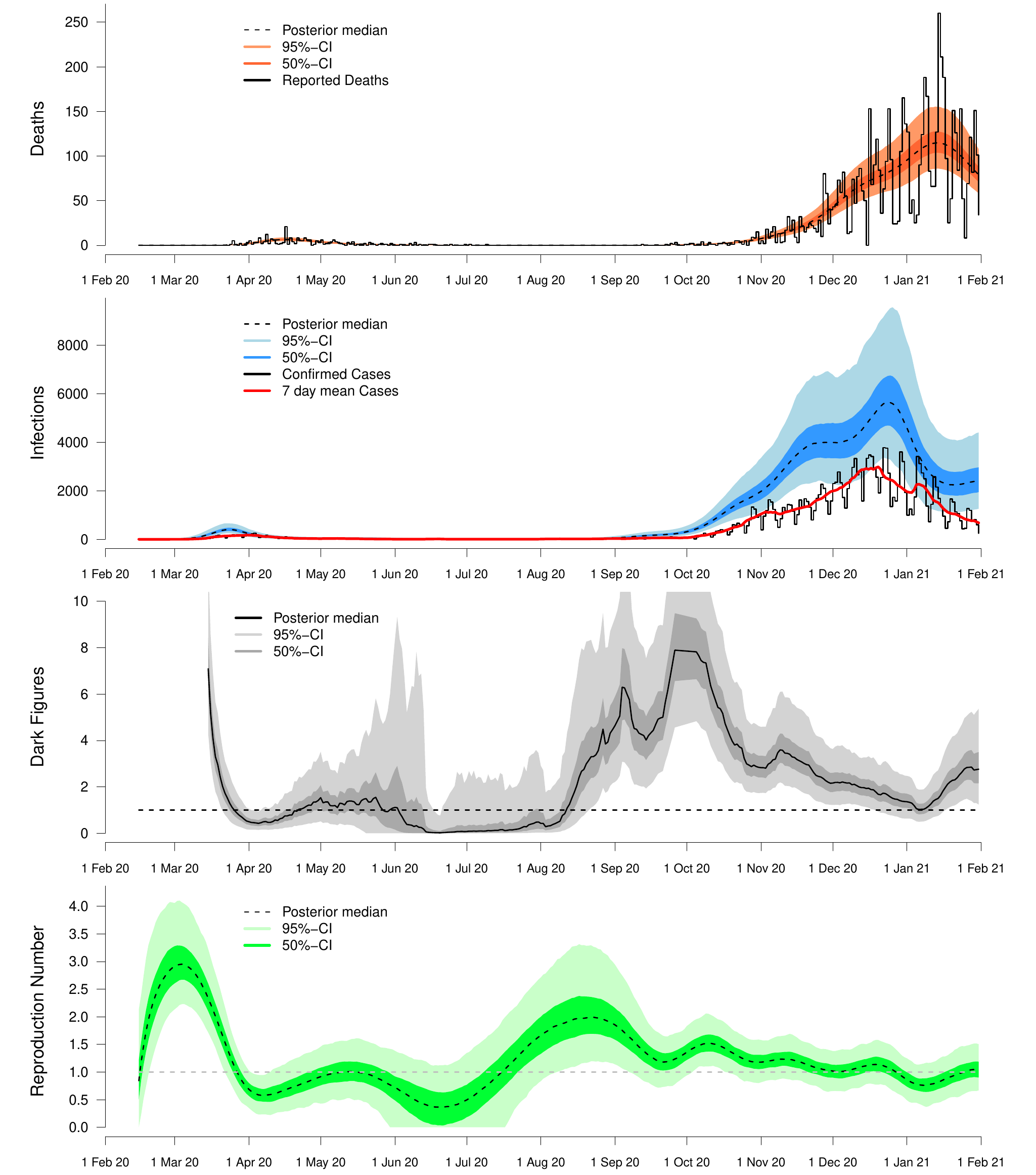}
 \caption{Results for Saxony based on age-specific IFR estimates from Brazeau et al.~\cite{brazeau2020}.}
\end{figure}

\clearpage

\subsubsection{Saxony-Anhalt}

\begin{figure}[h]
 \centering
 \includegraphics[width=\textwidth]{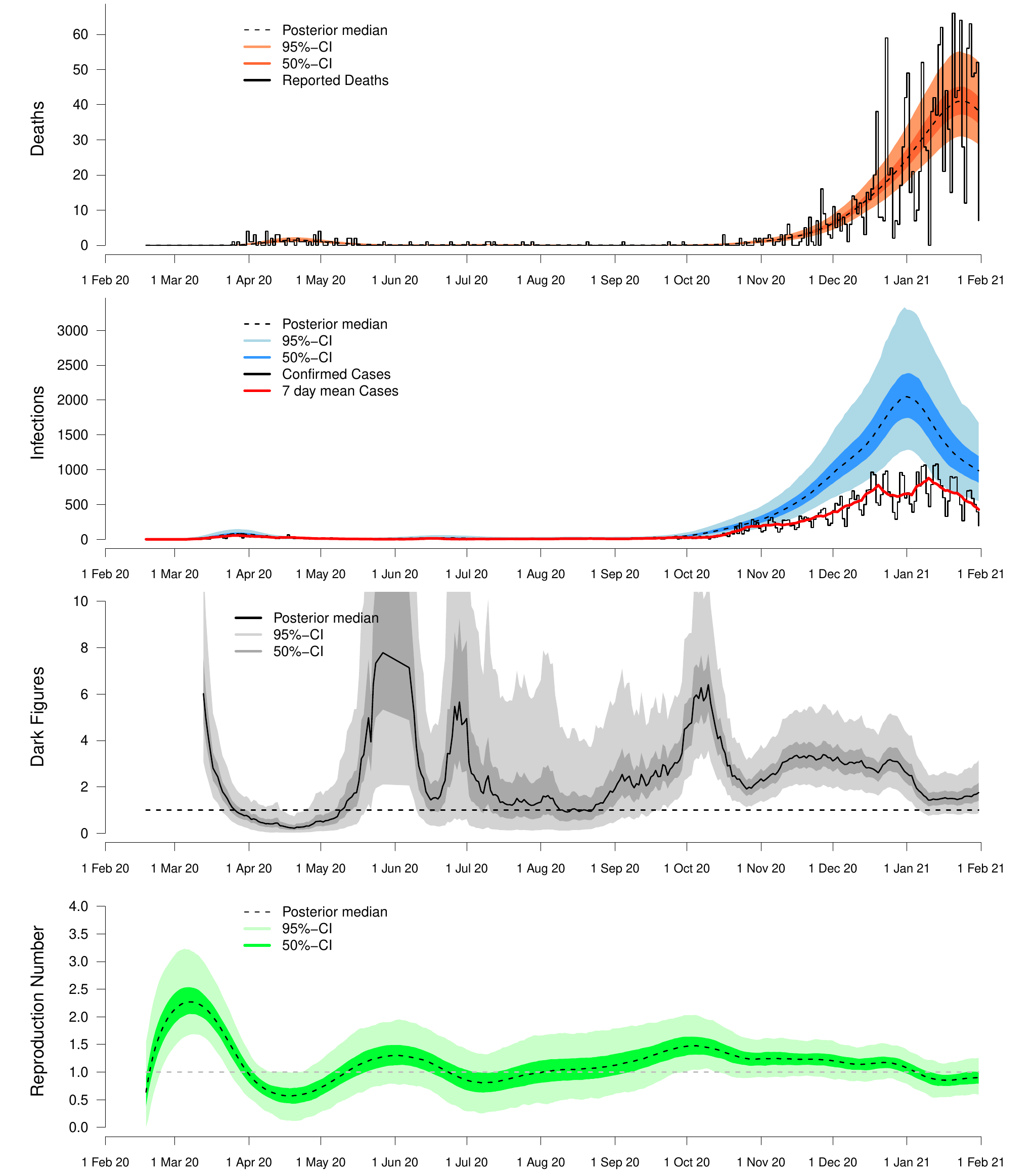}
 \caption{Results for Saxony-Anhalt based on age-specific IFR estimates from Brazeau et al.~\cite{brazeau2020}.}
\end{figure}

\clearpage

\subsubsection{Schleswig-Holstein}

\begin{figure}[h]
 \centering
 \includegraphics[width=\textwidth]{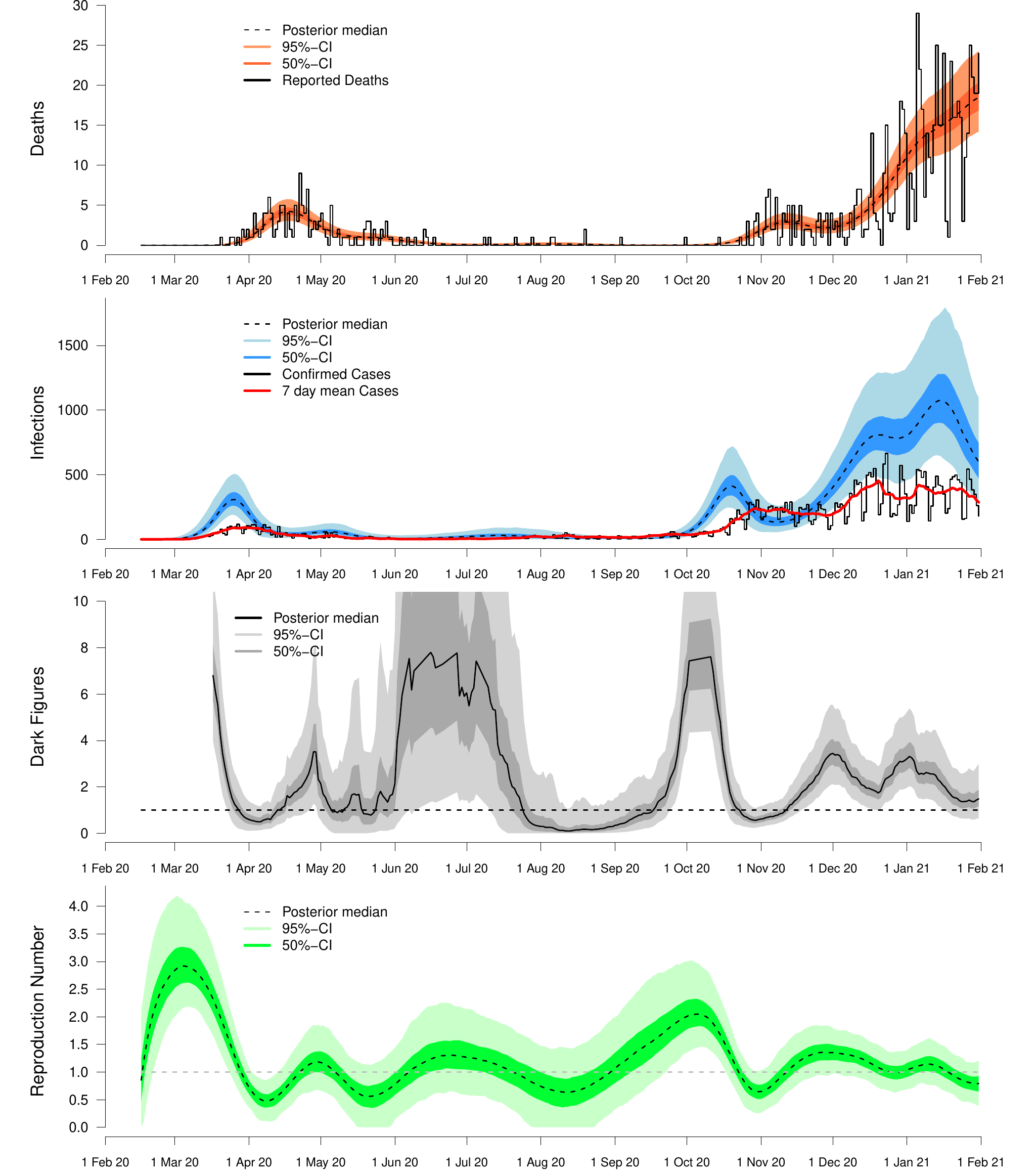}
 \caption{Results for Schleswig-Holstein based on age-specific IFR estimates from Brazeau et al.~\cite{brazeau2020}.}
\end{figure}

\clearpage

\subsubsection{Thuringia}

\begin{figure}[h]
 \centering
 \includegraphics[width=\textwidth]{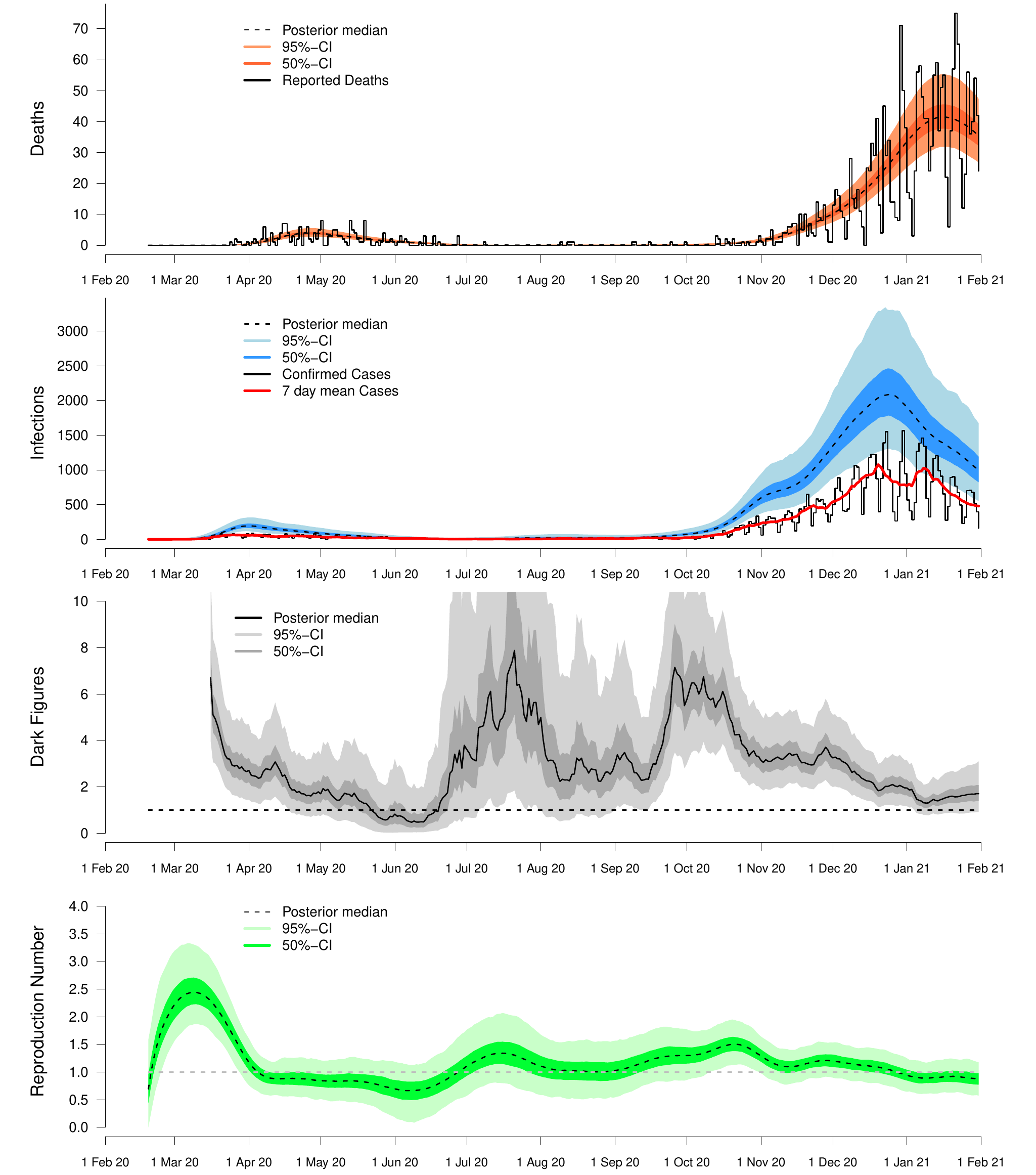}
 \caption{Results for Thuringia based on age-specific IFR estimates from Brazeau et al.~\cite{brazeau2020}.}
\end{figure}

\clearpage

\subsection{Diagnostic Plots for MCMC convergence} \label{ssec:diagnostics}

\begin{figure}[h]
 \centering
 \includegraphics[width=0.6\textwidth]{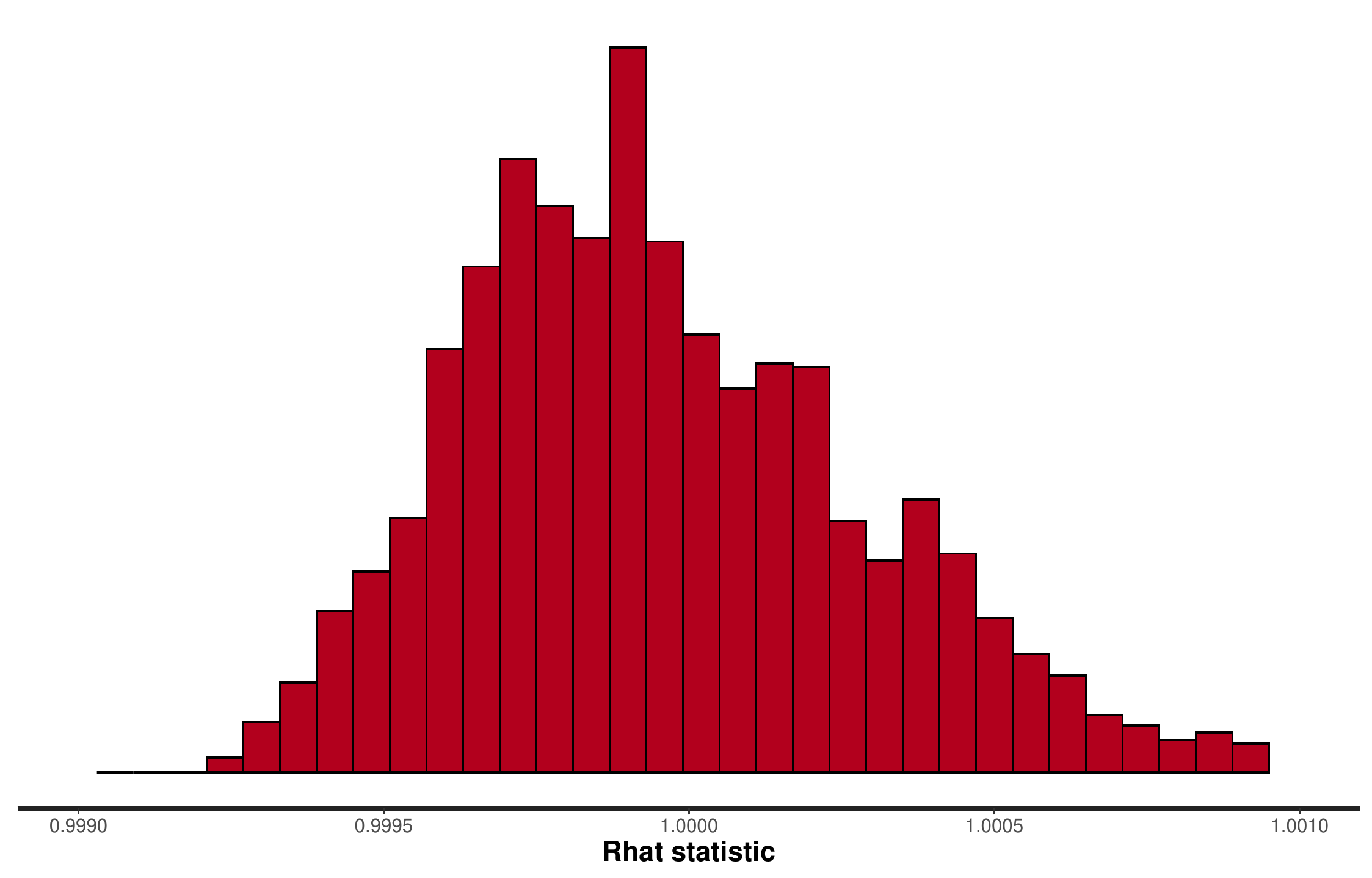}
 \caption{Histogram of Gelman-Rubin statistics~(Rhat) for all parameters of the hierarchical model for Germany (values below 1.1 and close to 1 are desirable).} 
\end{figure}

\begin{figure}[h]
 \centering
 \includegraphics[width=\textwidth]{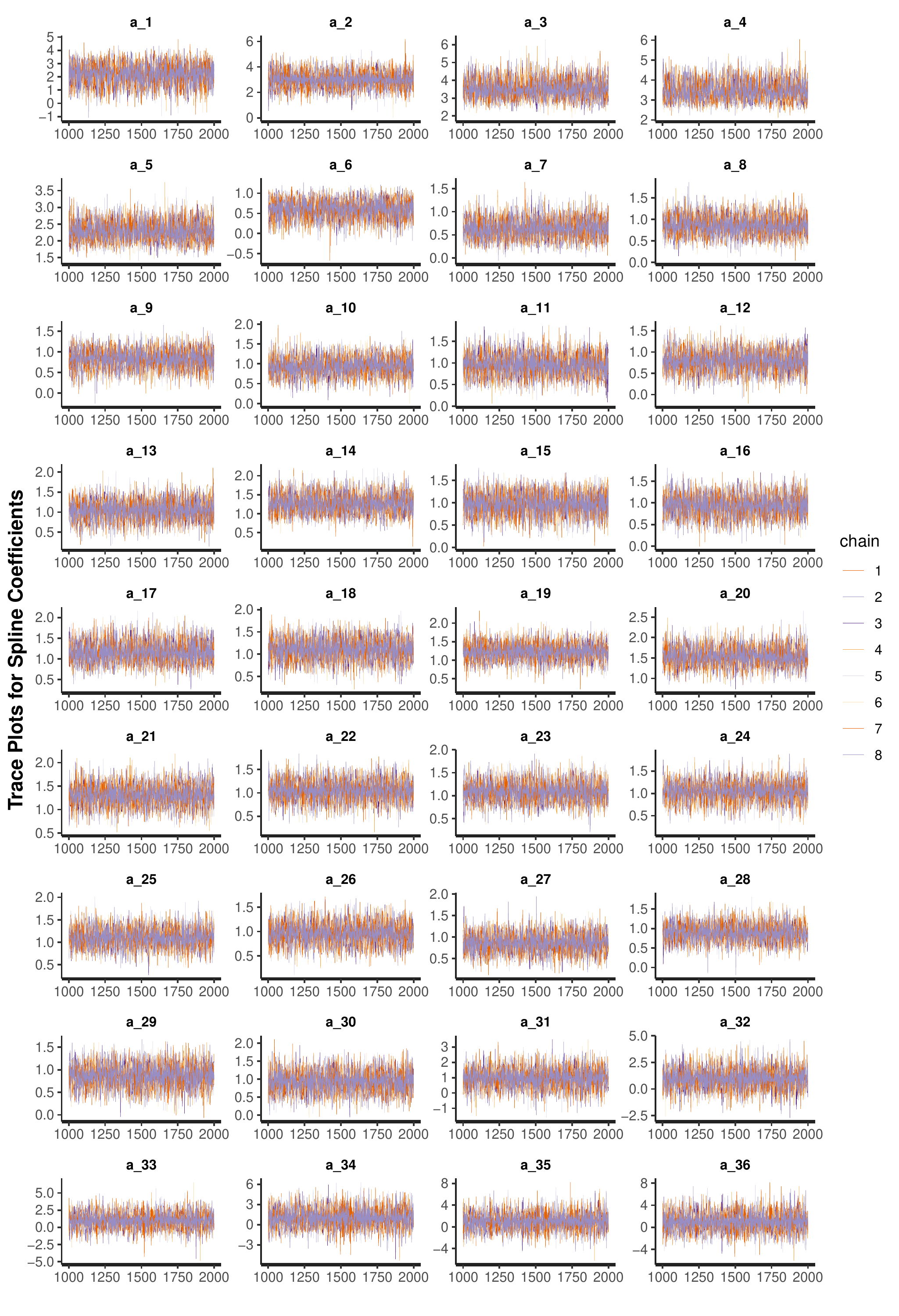}
 \caption{Trace plots of posterior samples for all spline coefficients of the hierarchical model for Germany. The eight independent chains appear to have mixed well. Trace plots for further model parameters show a similar mixing behaviour.}
\end{figure}

\begin{figure}[h]
 \centering
 \includegraphics[width=\textwidth]{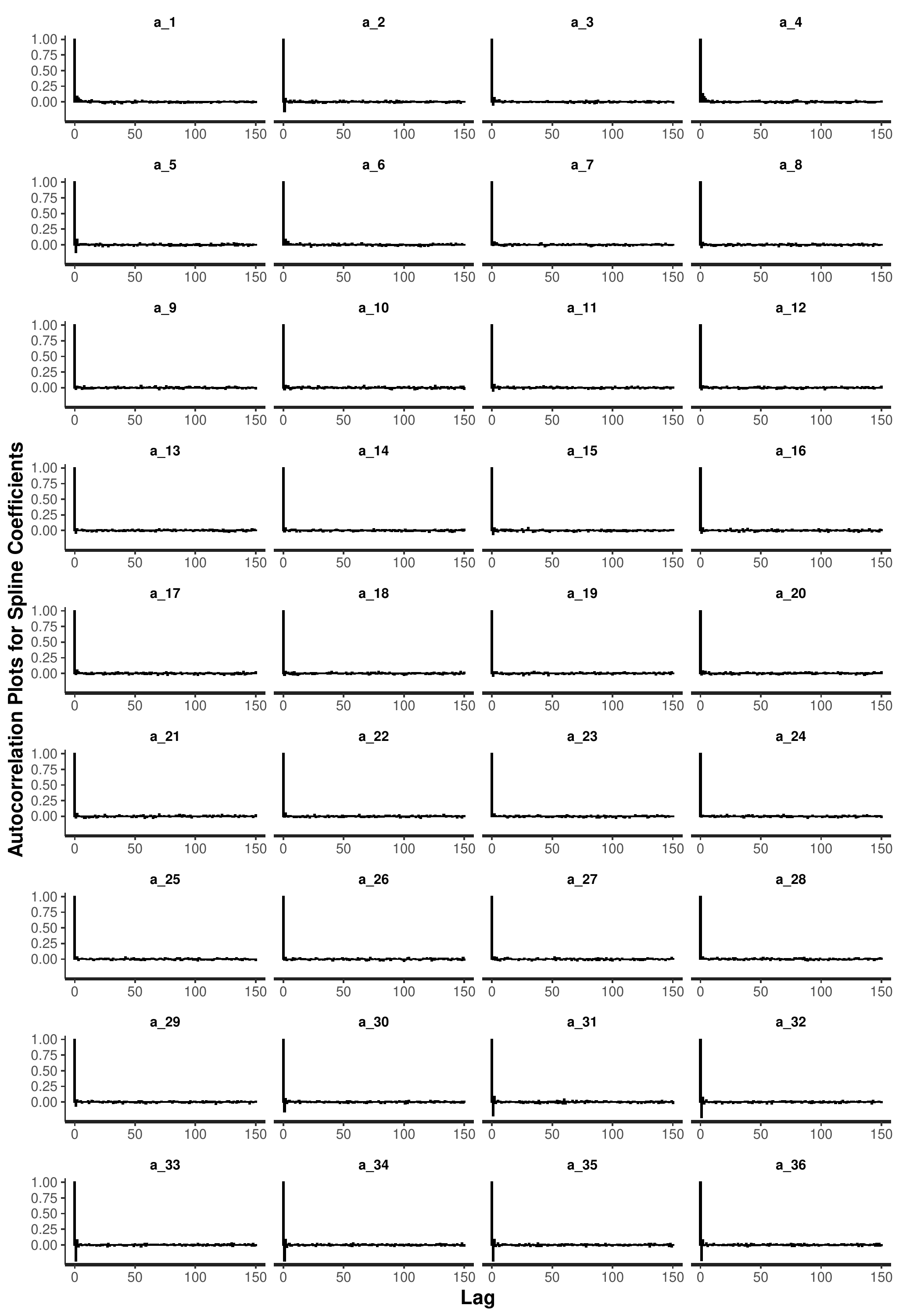}
 \caption{Autocorrelation plots for all spline coefficients of the hierarchical model for Germany, averaged over the eight independent chains. Autocorrelation plots for further model parameters are similar.}
\end{figure}

\end{document}